\documentclass[numsec,webpdf,contemporary,small,namedate]{oup-authoring-template}

\usepackage{subfig}
\usetikzlibrary{positioning, arrows.meta, angles, fit, quotes}

\theoremstyle{thmstyleone}%
\newtheorem{theorem}{Theorem}
\newtheorem{lemma}{Lemma}
\newtheorem{corollary}{Corollary}
\newtheorem{proposition}{Proposition}
\theoremstyle{thmstyletwo}%
\newtheorem{example}{Example}%
\newtheorem{remark}{Remark}%
\theoremstyle{thmstylethree}%

\newtheorem{assumption}{Assumption}

\begin{document}

\journaltitle{Journal Title Here}
\DOI{DOI HERE}
\copyrightyear{2022}
\pubyear{2019}
\access{Advance Access Publication Date: Day Month Year}
\appnotes{Paper}

\firstpage{1}


\title[FLGP]{Scalable Bayesian inference for heat kernel Gaussian processes on manifolds
} 
\author[1]{Junhui He}
\author[2]{Guoxuan Ma}
\author[2,$\ast$]{Jian Kang}
\author[1,$\ast$]{Ying Yang}

\authormark{He et al.}

\address[1]{\orgdiv{Department of Statistics and Data Science}, \orgname{Tsinghua University}, \orgaddress{\state{Beijing}, \country{China}}}
\address[2]{\orgdiv{Department of Biostatistics}, \orgname{University of Michigan, Ann Arbor}, \orgaddress{\state{Michigan}, \country{United States}}}

\corresp[$\ast$]{\emph{Address for correspondence:} Jian Kang (jiankang@umich.edu) and Ying Yang (yangying@tsinghua.edu.cn)}




\abstract{
We establish a scalable manifold learning method and theory, motivated by the problem of estimating fMRI activation manifolds in the Human Connectome Project (HCP). Our primary contribution is the development of an efficient estimation technique for heat kernel Gaussian processes in the exponential family model. This approach handles large sample sizes $n$, preserves the intrinsic geometry of data, and significantly reduces computational complexity from $\mathcal{O}(n^3)$ to $\mathcal{O}(n)$ via a novel reduced-rank approximation of the graph Laplacian's transition matrix and a Truncated Singular Value Decomposition for the eigenpair computation. The numerical experiments demonstrate the scalability and improved accuracy of our method for manifold learning tasks involving complex large-scale data.
}



\maketitle


\section{Introduction}

\begin{figure}
    \centering
    \includegraphics[width=0.6\textwidth]{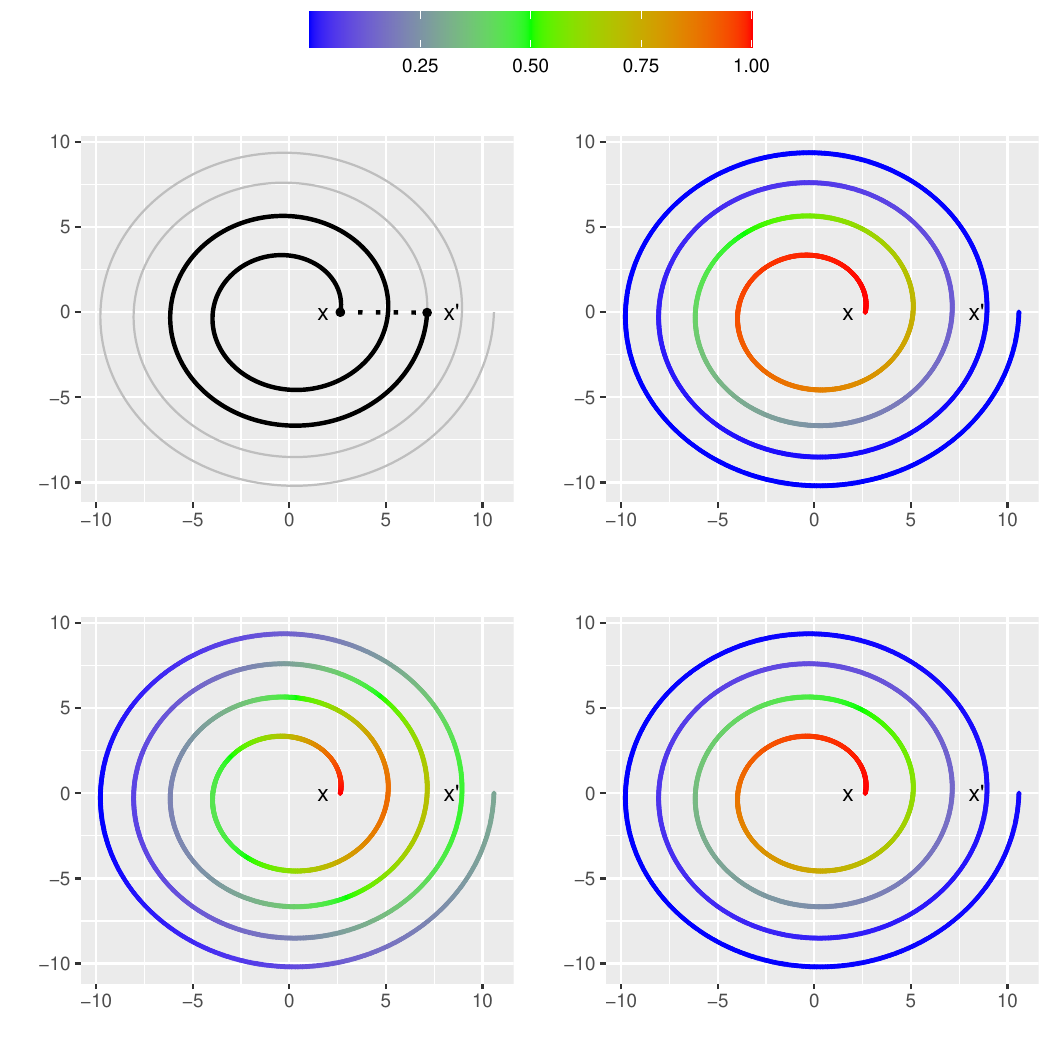}
    \caption{The non-Euclidean spiral case. Top left: distances on the spiral. Top right: heat kernels on $x$. Bottom left: squared exponential kernels on $x$. Bottom right: FLGP kernel estimates on $x$.}
    \label{fig:spiral}
\end{figure}

In recent years, the machine learning community has encountered a variety of complex data sets, such as image and network data, which often possess non-Euclidean geometry. This has contributed to a growing interest in manifold learning. To illustrate the non-Euclidean geometry, consider the example of a spiral curve, a one-dimensional submanifold of $\mathbb{R}^2$. As shown in the top left panel of Figure \ref{fig:spiral}, for two points $x,x'$ in the curve, the Euclidean distance is the length of the straight line segment connecting them (black dotted line), whereas the geodesic distance is the length of the spiral curve connecting them (black solid line). It is well known that many covariance kernels used in Gaussian process (GP) models rely on the distance, such as the squared exponential kernel and the Mat\'ern kernel~\citep{genton2001classes,rasmussen2006gaussian}. The posterior contraction rates in these models usually depend on the dimension of the data, as discussed in \citet{10.1214/009053607000000613,10.1214/08-AOS678}. Consequently, the differences between the Euclidean geometry and the intrinsic geometry substantially affect statistical inference.

In this article, we focus on the development of a domain-adaptive kernel by introducing a novel approach, the Fast Graph Laplacian Estimation for Heat Kernel Gaussian Processes (FLGP). To enhance clarity, we present a comparative visualization of intrinsic kernels and Euclidean-based kernels on the spiral in Figure \ref{fig:spiral}. The top right panel of the figure clearly demonstrates that the heat kernel respects the intrinsic geometry of the manifold. 
In contrast, the squared exponential kernel, depicted in the bottom left panel, is predicated on Euclidean distance. 
Our proposed FLGP method captures the intrinsic geometry of the manifold by achieving an efficient estimation of the heat kernel. For a more detailed quantitative analysis, we present kernel density plots in Section C of supplementary materials, which further illustrate the accuracy of various kernel estimates. 
Additionally, we have applied both FLGP and traditional Euclidean-based kernels (\emph{e.g.}, the squared exponential kernel) to make predictions on the spiral dataset. The results are compelling: FLGP yields a root mean square error (RMSE) of 0.343, whereas the Euclidean GP results in a significantly higher RMSE of 2.323. This comparison underscores the importance of incorporating the intrinsic geometry of the data when constructing GP models. Without accounting for the spiral's intrinsic structure, the Euclidean-based GP significantly underperforms in comparison to the domain-adaptive FLGP.

The manifold learning is also motivated by the analysis of biomedical imaging data.  Traditional analysis of volumetric functional Magnetic Resonance Imaging (fMRI) data, used in studying brain activation signals, often assumes that these signals exist within a regular three-dimensional Euclidean space. However, this assumption may not capture the complex nature of brain activation patterns. The human brain's cortical surface, where these signals are mapped, is not a simple three-dimensional structure but a complex, folded surface. This unique geometry challenges standard Euclidean-based analysis methods. Manifold learning methods fill this gap by offering a more sophisticated approach to modeling brain imaging data. They acknowledge and leverage the intricate geometry of the brain's surface, enabling a more accurate understanding of the brain's functional architecture. This approach has the potential to lead to discoveries in neuroscience and medicine. Thus, in machine learning, whether in understanding complex brain patterns or in simpler geometric structures like spheres, the utilization of the intrinsic geometry of data is crucial for optimal performance.

We focus on the prediction problem with GPs in this article. Suppose we have $n$ observations consisting of $m$ labelled data $\{(x_i,y_i)\}_{i=1}^m\subset \mathcal{X}\times \mathcal{Y}$ and $n-m$ unlabelled data $\{x_i\}_{i=m+1}^n\subset \mathcal{X}$. Define the point cloud as $X^n=\{x_i\}_{i=1}^n$. 
We are interested in the cases where $\mathcal{X}$ is a Riemannian manifold. The geodesic distance is completely different from the Euclidean distance, making it inappropriate to treat $\mathcal{X}$ as an Euclidean space. Our objective is to use $X^n$ to estimate unknown $\mathcal{X}$ and to devise a prediction algorithm that respects the intrinsic geometry of $\mathcal{X}$. In the context of GPs, this involves constructing a domain-adaptive kernel rather than relying on a kernel-based solely on extrinsic distance. A straightforward approach is to substitute the Euclidean distance with the geodesic distance.  For example, in squared exponential kernels, for any $x,x'\in \mathcal{X}$, consider the kernel function $k(x,x')=\exp\{-\rho(x,x')^2/2\epsilon^2\}$, where $\rho$ represents the geodesic distance and $\epsilon>0$. However, such kernels are generally not semi-positive definite and hence, do not qualify as covariance functions \citep{Feragen_2015_CVPR}. Our primary goal is to develop a valid domain-adaptive kernel for GPs on Riemannian manifolds.

Suppose 
$\mathcal{X}$ is a $d$-dimensional submanifold of $\mathbb{R}^p$. When $\mathcal{X}$ is not a submanifold of $\mathbb{R}^p$, such as Kendall shape spaces~\citep{kendall2009shape,https://doi.org/10.1111/rssb.12402} and Grassmann manifolds \citep{jmlee2012manifold}, we can still embed it into a high-dimensional Euclidean space by the Nash Embedding Theorem; 
see extrinsic GPs in \citet{lin2019extrinsic}. For such non-Euclidean domains, \citet{10.1214/15-AOS1390} use a simple squared exponential kernel Gaussian process prior with the intrinsic dimension adaptive random scaling parameter. They achieve a posterior contraction rate depending on the intrinsic dimension $d$. \citet{castillo2014thomas} construct a Gaussian process prior with the rescaling heat kernel on $\mathcal{X}$. They show that the heat kernel characterizes the geometry on $\mathcal{X}$ by the heat diffusion. The posterior distribution yields an optimal contraction rate adaptive to the manifold. However, $\mathcal{X}$ is unknown in many practical cases, and the heat kernel is analytically intractable except for some trivial cases such as Euclidean spaces or spheres. Thus, we have to calculate the heat kernel numerically.

Some researchers have leveraged the connection between the heat kernel and the transition density of Brownian Motions (BM) on manifolds to estimate the heat kernel. A primary challenge in this approach is simulating BM sample paths on $\mathcal{X}$. \citet{niu2019intrinsic} achieve this by solving systems of stochastic differential equations using the coordinate chart, while \citet{ye2020heat} utilize the exponential map for simulation. However, even if $\mathcal{X}$ is given, obtaining a global parameterization or an explicit exponential map is generally not feasible. Moreover, $\mathcal{X}$ remains unknown in many scenarios. To address this, \citet{niu2023intrinsic} employ probabilistic latent variable models to estimate the probabilistic parameterization and simulate BM trajectories on the unknown manifold using the probabilistic metric. 
The procedure of sampling trajectories is computationally expensive, as it necessitates the simulation of numerous sample paths originating at each input. Furthermore, the resulting kernel using trajectory-based methods is not inherently guaranteed to be semi-positive definite, and indeed often turns out not to be such. 

Furthermore, the spectral property suggests that the heat kernel can be constructed as a sum of eigenpairs of the Beltrami--Laplace operator on the manifold. \citet{10.1111/rssb.12486} estimate this operator using a Gaussian kernelized graph Laplacian on $X^n$, proposing a Graph Laplacian Gaussian Process (GLGP) for regression tasks. Their algorithm estimates the heat kernel by summing finitely many eigenpairs of the graph Laplacian. 
Similarly, \citet{NEURIPS2023_d611d06e} construct a graph Mat\'ern kernel relying on the eigenpairs of the normalized random walk graph Laplacian. 
Given that the graph Laplacian is an $n\times n$ matrix in these methods, the eigen-decomposition involves a time complexity of $\mathcal{O}(n^3)$, which becomes unaffordable for large-scale datasets.

Beyond the heat kernel, alternative methods that respect the geometry include embedding $\mathcal{X}$ into a lower-dimensional Euclidean space while preserving the intrinsic distance \citep{roweis2000nonlinear,belkin2001laplacian,weinberger2006unsupervised}. Inferring on the lower-dimensional representation can improve the performance. However, finding such an embedding is exceedingly challenging when $\mathcal{X}$ is unknown and complex.

In this article, we propose an efficient graph Laplacian approach to estimate the heat kernel Gaussian processes. Our novel insight lies in constructing a two-step random walk on $X^n$ through a transition decomposition strategy. Initially, we subsample $s$ induced points from the point cloud $X^n$, denoted by $I^s=\{u_i\}_{i=1}^s$. We then decompose the one-step movement from $X^n$ to $X^n$ into two steps: a jump from $X^n$ to $I^s$ and a subsequent jump from $I^s$ back to $X^n$. This strategy results in a reduced-rank approximation of the row stochastic transition matrix. Subsequently, we employ a Truncated Singular Value Decomposition (TSVD) to calculate the eigenpairs, which dramatically accelerates the computational process. Finally, the heat kernel covariance matrix is constructed from the eigenpairs based on the spectral property. The time complexity decreases remarkably from $\mathcal{O}(n^3)$ to $\mathcal{O}(n)$. The proposed algorithm is evaluated through simulations and real-data applications, outperforming other algorithms in all scenarios.

Generally speaking, FLGP is particularly effective in scenarios where the intrinsic geometry of the domain $\mathcal{X}$ differs from the extrinsic geometry of the ambient space $\mathbb{R}^p$. 
Our proposed method offers a rapid estimation of heat kernels, 
making domain-adaptive GPs feasible and efficient for large-scale data applications. 
We implement the algorithms within an \texttt{R} package, leveraging the power of \texttt{Rcpp} to provide a seamless and convenient interface for \texttt{R} users. Furthermore, we develop a comprehensive theoretical framework for FLGP in the exponential family. 
The approximation error of heat kernels within the FLGP framework is meticulously assessed through the spectral convergence property of the graph Laplacian, complemented by a rigorous perturbation analysis. Subsequently, the predictive error bound is established based on the approximation covariance error, thereby ensuring a robust predictive performance of our model. Ultimately, we successfully derive the posterior contraction rates for heat kernel GPs in the exponential family. A pivotal strength of our approach is that the posterior contraction rates depend on the intrinsic dimension $d$, rather than the ambient dimension $p$. 

The article is organized as follows. Section 2 gives a preliminary about the exponential family with canonical parameterization and heat kernels on manifolds. Section 3 provides a detailed description of the proposed methodology, which consists of subsampling induced points, constructing a two-step random walk, solving TSVD, and computing the heat kernel estimate. Section 4 presents the main theoretical results. Section 5 illustrates the performance in various examples, and Section 6 discusses the future work.

The FLGP package is accessible in \url{https://github.com/junhuihe2000/FLGP}.
\section{Preliminary}

\subsection{Canonical exponential family}

Consider a pair of random variables $(X,Y)\in \mathcal{X}\times \mathcal{Y}$. Suppose that the conditional distribution of $Y$ given $X=x$ follows an exponential family distribution with a canonical parameterization \citep{shao2003mathematical}. Specifically, given a dominant measure $\mu$, the conditional probability density $\pi_\theta(y|x)$ takes the general form,
\begin{equation}
    \pi_\theta(y|x)=h(y)\exp{\left\{\theta(x) \kappa(y)-J(\theta(x))\right\}},
\label{eq:exp}
\end{equation}
where $h(y)\geq 0$ and $\kappa(y)$ are functions that depend only on $y$, $\theta$ is a function mapping from $\mathcal{X}$ to $\Theta$, $\Theta$ is a subset of an Euclidean space, and $J(\theta(x))=\log \big\{\int_{\mathcal{Y}} h(y)\exp\{\theta(x) \kappa(y)\}d\mu(y) \big\}$. This form of the exponential family is referred to as the Canonical Exponential Family (CEF), which we denote as $Y|X=x,\theta~\sim\text{CEF}(\theta(x))$. The exponential family includes many common models, such as Gaussian and Bernoulli distributions, and emerges in various fields of statistics and machine learning \citep{doi:10.1080/01621459.2019.1677471,10.1214/19-AOS1810}. 

Furthermore, a fundamental property of the exponential family is that the conditional expectation of the sufficient statistic $\kappa(Y)$ is given by $\mathbb{E}_{\theta}\{\kappa(Y)|X=x\} = J'(\theta(x))$, where $J'(\cdot)$ is the first derivative of $J(\cdot)$. This expectation serves as the quantity of interest in this article.

\subsection{Heat kernels on manifolds}

Let $\mathcal{X}$ be a $d$-dimensional Riemannian manifold equipped with the geodesic distance $\rho$. Assume that $G$ is a probability measure on $\mathcal{X}$ with a smooth, positive density function $g$ under the volume form $V$. The weighted Beltrami--Laplace operator on $(\mathcal{X},G)$ is denoted by $\Delta_x$ and the Dirichlet--Laplace operator is defined as $\mathcal{L} = -\Delta_x$. By construction, $\mathcal{L}$ is a non-negative definite, self-adjoint operator in the Hilbert space $L^2(\mathcal{X},G)$. 
The heat kernel $p_t(x,x')$  is the unique fundamental solution of the heat equation on the manifold $(\mathcal{X},G)$, satisfying the following conditions,
\begin{equation*}
    \begin{cases}
        \frac{\partial }{\partial t} p_t(x,x')=\Delta_x p_t(x,x'),\quad t\in \mathbb{R}^+,~x,x'\in \mathcal{X}, \\
        \lim_{t\to 0}p_t(x,x')=\delta(x,x'), \quad x,x'\in \mathcal{X},\\
    \end{cases}
\end{equation*}
where $\delta(x,x')$ is the Dirac delta function, and $t\in \mathbb{R}^+$ is the diffusion time. The heat kernel characterizes the diffusion of heat from $x$ to $x'$ on the manifold over time $t$.

Assuming that $\mathcal{X}$ is compact, the spectral theorem guarantees that the spectrum of $\mathcal{L}$ consists of a non-decreasing non-negative sequence $\{\lambda_i\}_{i=1}^\infty$ with $\lambda_1=0$ and $\lim_{i\to \infty} \lambda_i = +\infty$. There exists an orthonormal basis $\{e_i\}_{i=1}^\infty$ in $L^2(\mathcal{X},G)$ such that for $i\geq 1$, $e_i$ is the eigenfunction of $\mathcal{L}$ corresponding to $\lambda_i$. Consequently, for $t\in \mathbb{R}^+$ and any two points $x,x'\in\mathcal{X}$, $p_t(x,x')$ admits the spectral expansion:
\begin{equation}
\label{eq:spec}
    p_t(x,x') = \sum_{i=1}^\infty \exp(-t\lambda_i) e_i(x)e_i(x').
\end{equation}
If the eigenpairs of $\mathcal{L}$ can be accurately estimated, this expansion provides an effective method of approximating the heat kernel. Therefore, our objective is to develop a fast numerical method for estimating the eigenpairs of $\mathcal{L}$.
We refer readers to \citet{grigor2006heat,grigoryan2009heat} for further discussions of heat kernels on manifolds.
\section{Methodology}

Let the observed data comprise labeled observations $\{(x_i,y_i)\}_{i=1}^m$ and unlabeled observations $\{x_i\}_{i=m+1}^n$, which are independent and identically distributed (\emph{i.i.d.}) samples drawn from the population distribution $(X,Y)\in \mathcal{X}\times \mathcal{Y}$. Define the sets of labeled inputs and outputs as $X^m=\{x_i\}_{i=1}^m$ and $Y^m=\{y_i\}_{i=1}^m$, respectively.  Assume that the input variable $X$ follows a distribution $G$ that is absolutely continuous with respect to the volume form on $\mathcal{X}$, and denote the marginal density as $g(x)$. For any $x\in\mathcal{X}$, let the conditional distribution of $Y$ be given by $\pi_\theta(y|x) \sim \text{CEF}\{\theta(x)\}$, where $\theta:\mathcal{X}\rightarrow \Theta$ is an unknown function. 

Consider the mean function $\varphi(x)= \mathbb{E}_\theta\{\kappa(Y)|X=x\}$, where the state space of $\varphi$ depends on the specific likelihood and may not necessarily be identical to $\mathbb{R}$. We assume a relationship between $\varphi$ and a latent function $f$ through a well-chosen link function $l$, such that $\varphi = l\circ f$. To induce a prior on $\varphi$, we assign a heat kernel Gaussian process to $f$. Specifically, we assume
\begin{equation}
\label{eq:GP}
    f\mid t\sim \mathcal{GP}\{0,p_t(x,x')\},
\end{equation}
where $t>0$ is the diffusion time, and $p_t(x,x')$ is the heat kernel on $\mathcal{X}$, as defined in $\eqref{eq:spec}$.

Let $f^m$ be the $m$-dimensional vector consisting of the values $\{f(x_i)\}_{i=1}^m$. The posterior expectation of the mean function in a test input $x^*$ is given by
\begin{equation*}
    \mathbb{E}[\varphi^*|X^{m},Y^{m},x^*] = \int\int l(f^*)\pi(f^*|f^m,x^*)\pi(f^m|X^m,Y^m) df^m df^*,
\end{equation*}
where $\pi(f^m|X^m,Y^m)=\pi(Y^m|f^m)\pi(f^m|X^m)/\pi(Y^m|X^m)$ is the posterior density over the latent variables. Here, $\varphi^*$ and $f^*$ correspond to $\varphi(x^*)$ and $f(x^*)$, respectively.

\subsection{Fast graph Laplacian heat kernel estimation}\label{sec:FGLHKE}

When dealing with a Riemannian manifold $\mathcal{X}$, traditional Euclidean kernels like the squared exponential kernel often result in suboptimal performance because they do not leverage the manifold's geometry. To address this, it is crucial to develop a domain-adaptive kernel that respects the manifold's intrinsic geometry. The heat kernel $p_t$, derived from the diffusion process on the manifold, naturally incorporates this geometric information \citep{coifman2006diffusion}. Furthermore, as a semi-positive definite kernel, $p_t$ is well-suited as a covariance function for Gaussian processes.

From an intuitive standpoint, the heat kernel $p_t$ can be seen as a generalization of the squared exponential kernel to the context of Riemannian manifolds, inheriting many of its beneficial properties. This makes the heat kernel Gaussian process an excellent alternative for domain-adaptive GPs. 
However, the heat kernel's analytical intractability, except in special cases such as spheres and homogeneous spaces of compact Lie groups \citep{azangulov2023stationarykernelsgaussianprocesses}, presents a challenge, even with known $\mathcal{X}$. 

In this article, we establish a fast graph Laplacian approach to estimate the heat kernel efficiently, enabling the construction of a heat kernel GP in large-scale datasets where $\mathcal{X}$ is unknown. Our method aims to approximate the heat kernel covariance matrix with both labeled observations $\{(x_i,y_i)\}_{i=1}^m$ and unlabeled observations $\{x_i\}_{i=m+1}^n$, all while minimizing computational cost. 

Additionally, we introduce a convenient notation for summing matrix elements: for any matrix $B\in \mathbb{R}^{l\times q}$, we define $B_{i\cdot}=\sum_{j=1}^q B_{ij}$ as the sum of the $i$-th row, and $B_{\cdot j}=\sum_{i=1}^l B_{ij}$ as the sum of the $j$-th column, for $1\leq i \leq l$ and $1\leq j \leq q$. We now detail the Fast Graph Laplacian Estimation for Heat Kernel Gaussian Process (FLGP) algorithm in a step-by-step manner as follows.

Step 1, subsample induced points. We begin by subsampling $s$ induced points from the point cloud $X^n$, denoted as $I^s = \{u_i\}_{i=1}^s$. This induced point set is used to simplify $X^n$ and to represent the domain geometry. In this article, we consider two subsampling methods: k-means clustering and random sampling.

Step 2, choose a base kernel. We select a suitable kernel function $k(x,x')$, such as the squared exponential kernel, defined as $k(x,x') = \exp\left(-||x-x'||^2/4\epsilon^2\right)$, where $\epsilon>0$ is the bandwidth. We construct the cross kernel matrix $K$ between $X^n$ and $I^s$ with elements $K_{ij}=k(x_i,u_j)$ for $1\leq i\leq n$ and $1\leq j \leq s$. The resulting $K$ is an $n\times s$ matrix.

Step 3, construct cross similarity and transition matrices. We define $A$ and $Z$ as the cross similarity and transition matrices, respectively. For $1\leq j \leq s$, let $n_j$ be the number of points in $X^n$ whose nearest induced point is $u_j$. Specifically, in k-means clustering, $\{n_j\}_{j=1}^s$ corresponds to the size of each cluster. The elements of $A$ are given by,
\begin{equation}
\label{eq:W}
    A_{ij}= \frac{n_jk(x_i,u_j)}{\sum_{q=1}^n k(x_q,u_j)\sum_{q=1}^s n_q k(x_i,u_q)},
\end{equation}
for $1\leq i\leq n$ and $1\leq j \leq s$. The work of \citet{coifman2006diffusion} suggests that this normalization is to adjust for the non-uniform sampling density. The cross transition matrix $Z$ is calculated as $Z_{ij}=A_{ij}/A_{i\cdot}$ for $1\leq i\leq n$ and $1\leq j \leq s$. 
Notably, $Z$ can serve as a probability transition matrix of random walks from $X^n$ to $I^s$.

Step 4, define graph Laplacian. Let $\Lambda$ be a diagonal matrix such that $\Lambda_{jj} = Z_{\cdot j}$ for $1\leq j \leq s$. We define the graph Laplacian $L$ associated with subsampling as
\begin{equation}
\label{eq:gl}
    L = I - (Z\Lambda^{-1} Z^\top)^{1/2}.
\end{equation}
We perform a Truncated Singular Value Decomposition (TSVD) on $Z\Lambda^{-1/2}$ to obtain the singular values $\{\sigma_{i,\epsilon}\}_{i=1}^s$ and the $l^2$-normalized left singular vectors $\{v_{i,\epsilon}\}_{i=1}^s$. Let $\lambda_{i,\epsilon} = 1-\sigma_{i,\epsilon}$ for $1\leq i\leq s$, and thus the eigenpairs of $L$ are $\{(\lambda_{i,\epsilon},v_{i,\epsilon})\}_{i=1}^s$. Suppose $0=\lambda_{1,\epsilon}\leq \cdots\leq \lambda_{s,\epsilon}\leq 1$.

Step 5, estimate the heat kernel. We choose the smallest $M$ eigenvalues $\{\lambda_{i,\epsilon}\}_{i=1}^M$ and corresponding eigenvectors $\{v_{i,\epsilon}\}_{i=1}^M$ of $L$. The heat kernel is estimated based on Equation \eqref{eq:spec} as follows,
\begin{equation}
\label{eq:hkcov}
    C_{\epsilon,M,t} = n\sum_{i=1}^M\exp(-t\lambda_{i,\epsilon}/\epsilon^2) v_{i,\epsilon}v_{i,\epsilon}^\top,
\end{equation}
where $t>0$ is the diffusion time.

\begin{figure}
    \centering
    \resizebox{0.8\textwidth}{!}{%
    \begin{tikzpicture}[every node/.style={align=center},
point/.style={draw,circle,minimum size=2.5em,inner sep=1},
node distance = 5mm and  6.5mm]

\node(cloud) [point,draw opacity=0] {$X^n:$};
\node(x1) [right=of cloud] [point] {$x_1$};
\node(x2) [right=of x1] [point] {$x_2$};
\node(x3) [right=of x2] {$\cdots$};
\node(x4) [right=of x3] [point] {$x_{n-1}$};
\node(x5) [right=of x4] [point] {$x_{n}$};

\node(u1) [below=of x2] [point] {$u_1$};
\node(u2) [right=of u1] {$\cdots$};
\node(u3) [right=of u2] [point] {$u_s$};
\node(tem) [left=of u1] [point, opacity=0] {};
\node(induced) [left=of tem] [point,draw opacity=0] {$I^s:$};

\node(x-2) [below=of u1] [point] {$x_2$};
\node(x-1) [left=of x-2] [point] {$x_1$};
\node(x-3) [right=of x-2] {$\cdots$};
\node(x-4) [right=of x-3] [point] {$x_{n-1}$};
\node(x-5) [right=of x-4] [point] {$x_n$};
\node(cloud-again) [left=of x-1] [point,draw opacity=0] {$X^n:$};

\node(rw) [draw, dashed, fit={(cloud) (x5) (cloud-again) (x-5)}] {};

\node(A3) [left=of rw] [draw] {Construct the cross \\ transition matrix $Z$};
\node(A2) [above=of A3] [draw] {Calculate the cross \\ kernel matrix $K$};
\node(A1) [above=of A2] [draw] {Subsample $I^s$};
\node(A4) [below=of A3] [draw] {Do TSVD  for $Z\Lambda^{-1/2}$};
\node(A5) [below=of A4] [draw] {Estimate the heat \\ kernel $C_{\epsilon,M,t}$};

\node [draw, thick, fit={(A1) (A2) (A3) (A4) (A5) (rw)}] {};

\draw[arrows = -Stealth]
(x1) edge (u1) (x1) edge (u3)
(x2) edge (u1) (x2) edge (u3)
(x4) edge (u1) (x4) edge (u3)
(x5) edge (u1) (x5) edge (u3)
;

\draw[arrows = -Stealth]
(u1) edge (x-1) (u1) edge (x-2) (u1) edge (x-4) (u1) edge (x-5)
(u3) edge (x-1) (u3) edge (x-2) (u3) edge (x-4) (u3) edge (x-5)
;

\draw[arrows = -Stealth]
(cloud) edge["$Z$"] (induced)
(induced) edge["$\Lambda^{-1}Z^\top$"] (cloud-again)
;

\draw[arrows = -Stealth]
(A1) edge (A2)
(A2) edge (A3)
(A3) edge (A4)
(A4) edge (A5);

\draw[dashed] (A3) edge (rw);

\end{tikzpicture}
}
    \caption{The flow chart of FLGP to estimate the heat kernel. The transition decomposition strategy contributes to the construction of a reduced-rank stochastic matrix in random walks.}
    \label{fig:FLGP}
\end{figure}

From a graph-based perspective, we can consider a bipartite graph \citep{diestel2024graph} $\mathcal{G}=(\mathcal{V},\mathcal{U},\mathcal{E})$ with the sets $\mathcal{V}$ and $\mathcal{U}$ representing the two disjoint vertex sets, where $\mathcal{V}=\{x_i\}_{i=1}^n$, $\mathcal{U}=\{u_j\}_{j=1}^s$. The edge set $\mathcal{E}$ comprises edges connecting vertices in $\mathcal{V}$ to those in $\mathcal{U}$. The weight of the edge between $x_i$ and $u_j$ is designated as $Z_{ij}$, resulting in a biadjacency matrix $Z$. Then the full adjacency matrix for the bipartite graph is $\begin{pmatrix}0&Z\\Z^\top&0 \end{pmatrix}$, where $Z^\top$ is the transpose of $Z$. Upon this bipartite graph $\mathcal{G}$, we establish random walks by defining the one-step transition probability matrices as $Z$ and $\Lambda^{-1}Z^\top$. Consequently, $L$ can be interpreted as the two-step random walk graph Laplacian on the graph $\mathcal{G}$. 

The whole procedure of FLGP is shown in Figure \ref{fig:FLGP}. After training, we can make inference in GPs with the estimated covariance matrix $C_{\epsilon,M,t}$. Owing to the construction, we have the following proposition for the spectrum of $L$. 

\begin{proposition}
\label{prop:spectrum}
The eigenvalues of graph Laplacian $L$ in \eqref{eq:gl} are real-valued and lie in $[0,1]$. In particular, $0$ is the smallest eigenvalue of $L$.
\end{proposition} 

The key novelty in FLGP is to construct a reduced-rank approximation for the transition matrix. Then the eigendecomposition of $L$ is replaced by the TSVD of $Z\Lambda^{-1/2}$. We explain the core idea from the random walk viewpoint. According to the graph spectral theory, there is a close relationship between the heat kernel on manifolds and random walks on the point cloud $X^n$. 
Based on the induced points $I^s$, we decompose the direct random walk into two steps. For $x,x'\in X^n$, the movement from $x$ to $x'$ is decomposed into a jump from $x$ to $I^s$ and a jump from $I^s$ to $x'$. Then the transition probability is
\begin{equation*}
    q(x,x')=q(x,I^s)q(I^s,x') = \sum_{j=1}^s q(x,u_j)q(u_j,x').
\end{equation*}
Let the cross transition matrices be $Z$ and $\Lambda^{-1}Z^\top$, respectively. Then the two-step transition probability matrix is $Z\Lambda^{-1}Z^\top$, yielding a reduced-rank approximation for the full-rank one-step transition matrix of \citet{10.1111/rssb.12486}.

The diagram inside the dashed rectangle in Figure \ref{fig:FLGP} illustrates the transition decomposition graphically. As $I^s$ contains only $s$ induced points, the rank of $Z\Lambda^{-1}Z^\top$ is at most $s$, which is much less than $n$, the rank of the transition matrix in one-step random walks. 
Consequently, the complexity of the TSVD is $\mathcal{O}(ns^2+s^3)$ instead of $\mathcal{O}(n^3)$. This accelerates the computational speed in large-scale data. Sparsity will be introduced into the graph to remove the $ns^2$ term further.

\subsection{Sparsity and likelihood-free kernels}
Let $r$ be a positive integer. For each $x_i\in X^n$, denote $\{i_j\}_{j=1}^r$ as the indices of the $r$ closest induced points to $x_i$. The local induced point set of $x_i$ is defined as $\{u_{j}\}_{j\in\{i_1,\ldots, i_r\}}$. We introduce a sparse base kernel matrix by $K^*$, where for $1\leq i \leq n$, the element $K^*_{ij} = K_{ij}$ if $j\in \{i_1,\ldots,i_r\}$, and $K^*_{ij}=0$ otherwise. Thus, $K^*$ is an $n\times s$ sparse matrix with $nr$ non-zero elements. This sparsity in $K^*$ results in sparse cross similarity and transition matrices, each containing $nr$ non-zero elements, thereby saving the memory usage. Furthermore, the time complexity of the TSVD is reduced to $\mathcal{O}(nr^2+s^3)$. This sparsification technique significantly accelerates the computational process, especially when $r<<s$. Additionally, as shown by \citet{garcia2020error}, the sparsity may improve the accuracy of heat kernel estimation by inducing local connectivity on the graph Laplacian.

In the case of the squared exponential kernel, optimizing the bandwidth $\epsilon$ necessitates repeated and prohibitive computation of eigenpairs during the calculation of the marginal likelihood. Moreover, the likelihood is based solely on labeled observations, thereby not fully leveraging the substantial information contained in unlabeled observations. To address these limitations, we propose the development of a likelihood-free base kernel, which is constructed exclusively from the point cloud, independent of the likelihood. In our work, we adopt the Local Anchor Embedding (LAE) kernel, as introduced by \citet{liu2010large}. The kernel matrix, denoted as $K^\mathrm{lae}$, is calculated by minimizing the square loss of a reconstruction problem. Specifically, for $1\leq i \leq n$, we have
\begin{equation*}
    \begin{split}
        &\min \left|\left|x_i-\sum_{j\in \{i_1,\ldots,i_r\}}K^\mathrm{lae}_{ij}u_{j}\right|\right|^2,\\
        & \emph{s.t.}\quad  K^\mathrm{lae}_{ij}\geq0\text{~for~}j\in \{i_1,\ldots,i_r\}, ~ \sum_{j \in \{i_1,\ldots,i_r\}}K^\mathrm{lae}_{ij}=1.\\
    \end{split}
\end{equation*}
This task is a convex optimization problem, which is amenable to solution by the Nesterov gradient method \citep{nesterov2018lectures}. The optimal solution corresponds to the coefficients of the projection onto a convex hull, with all other elements being set to zero. 
\subsection{Algorithms and the complexity analysis}

Given the covariance $\mathcal{C}$, we can calculate the marginal likelihood of GPs based on the concrete models as follows,
\begin{equation*}
    \pi(y|x,\mathcal{C},\sigma_\eta)=\int \pi(y|f,x,\sigma_\eta)\pi(f|\mathcal{C},x)df,
\end{equation*}
where $\sigma_\eta$ is the optional and potential nuisance parameter, such as noise variance in GP regression. 
The parameters of FLGP can be categorized into three groups:
\begin{enumerate}
    \item Hyperparameters: These include variables such as $s,r$ and $M$, which are predefined by the user. In Section E of supplementary materials, we conduct a sensitivity analysis to systematically assess how variations in these hyperparameters influence the model's predictive performance.
    \item Differentiable parameters: Parameters like $\sigma_\eta$ (if any) and $t$ fall into this category. The marginal likelihood is differentiable with respect to these parameters, enabling the application of gradient-based optimization techniques.
    \item Non-differentiable parameters: Parameters such as the bandwidth $\epsilon$ in squared exponential kernels are non-differentiable. The spectral decomposition of the graph Laplacian poses a challenge for differentiation, leading to prohibitive gradient calculations. Although \citet{NEURIPS2023_d611d06e} proposed a fully differentiable approach for efficiently training Mat\'ern Gaussian processes on implicit manifolds, this method is not applicable to our scenario. The sparse precision matrix structure of Mat\'ern kernels is not preserved in heat kernels. Consequently, we resort to the grid-search method for optimizing these non-differentiable parameters in this article.
\end{enumerate}
Here we present a concise overview of the Fast Graph Laplacian Estimation for Heat Kernel Gaussian Processes (FLGP) in Algorithm \ref{alg:FLGP}.

\begin{algorithm}
\small
\caption{FLGP algorithms on manifolds} \label{alg:FLGP}
\begin{algorithmic}[1]
\Statex \textbf{Input:} Labeled observations $\{(x_i,y_i)\}_{i=1}^m$, unlabeled observations $\{x_i\}_{i=m+1}^n$ and other hyperparameters $s,r,M$.
\State Subsample $s$ induced points $I^s$ from $X^n$ by random sampling or k-means clustering.
\State Construct the $n\times s$ sparse matrix $K$ with $nr$ non-zero elements by the squared exponential kernel or the local anchor embedding kernel.
\State Calculate the $n\times s$ sparse matrices $A$ and $Z$.
\State Compute the largest $M$ singular values and left singular vectors of $Z\Lambda^{-1/2}$. Obtain the corresponding eigenpairs of $L$.
\State Estimate the heat kernel covariance matrix as in \eqref{eq:hkcov}. \Comment{Only the $m\times m$ block in the upper left corner are required in the training.}
\State Optimize $\sigma_\eta$ (if any), $\epsilon$ (for the squared exponential kernel) and $t$ by maximizing the marginal likelihood on the labeled observations.
\Statex \textbf{Output:} The estimation of heat kernel Gaussian processes on the point cloud.
\end{algorithmic}
\end{algorithm}

\begin{proposition}
    The time complexity of the heat kernel estimate (training + test) by FLGP is $\mathcal{O}(nsp+nr^2+s^3+nmM)$ for random subsampling or $\mathcal{O}(nspI+nr^2+s^3+nmM)$ for k-means clustering, where $n$ is the number of all observations, $m$ is the number of labeled observations, $s$ is the number of induced points, $p$ is the ambient dimension, $r$ is the number of local induced points, $M$ is the number of eigenpairs used, and $I$ is the iteration number in k-means.
\end{proposition}

\begin{remark}
    The time complexity of FLGP is linearly dependent on $n$. Therefore, our algorithm allows scalable applications of heat kernel GPs. Conversely, the time complexity of GLGP is $\mathcal{O}(n^3)$. If $s<<n$, our approach is much faster than GLGP.
\end{remark}

In the proposed FLGP, the reduced-rank approximation of the graph Laplacian is constructed upon the graph with vertices corresponding to the point cloud $X^n$. This approach results in eigenvectors that are evaluated solely at the points of the original point cloud. Consequently, the heat kernel estimation is limited to the observed points $X^n$, and the kernel at a point entirely outside of the point cloud $X^n$ remains unknown. To overcome this limitation, we can explore interpolation techniques to extend the eigenvectors to new points, such as the Nystr\"om approximation in \citet{10.1111/rssb.12486} and \citet{NEURIPS2023_d611d06e}. We may incorporate an in-depth discussion of this problem in our future research. 

For the scope of this article, all numerical experiments are confined to inference within the original point cloud. Therefore, our proposed method suffices for the current cases, as it does not necessitate extrapolation beyond the provided data points. 
\section{Theory}

The theoretical framework consists of three key components. First, we establish error bounds for the numerical estimation of the heat kernel. This forms the foundation for our subsequent derivation of predictive error bounds. Finally, we prove the posterior contraction rate for GPs utilizing the \emph{exact} heat kernel on manifolds.

Let $\mathcal{X}$ be a domain manifold with geodesic distance denoted by $\rho$. Suppose $V$ represents the volume form on $\mathcal{X}$, and let $G$ be a probability measure on $\mathcal{X}$. For any $x\in\mathcal{X}$ and $t>0$, define the geodesic ball as $B_{t}(x)=\left\{z\in\mathcal{X}|~\rho(x,z)\leq t\right\}$.

\begin{assumption}
Suppose $\mathcal{X}$ is a compact, connected, $d$-dimensional manifold without boundary. Let $g(x)$ denote the smooth density function of $G$ with respect to the volume form $V$; see Chapter 2 in \citet{jmlee2012manifold} for details on smooth functions on manifolds. Assume there exist constants $g_{\min},g_{\max}>0$, such that $g_{\min} \leq g(x) \leq g_{\max}$ for all $ x\in \mathcal{X}$. Furthermore, suppose the point cloud $X^n$ consists of independently and identically samples drawn from the distribution $G$. We also assume the existence of positive constants $b_1,b_2,b_3,b_4>0$ such that, for $x,x'\in \mathcal{X}$, $t>0$, the heat kernel satisfies the following Gaussian estimates:
\begin{equation}
\label{eq:gaussianestimate}
    \frac{b_1}{F(x,x',t)}\exp{\left\{-\frac{b_2\rho^2(x,x')}{t}\right\}}\leq p_t(x,x')\leq \frac{b_3}{F(x,x',t)}\exp{\left\{-\frac{b_4\rho^2(x,x')}{t}\right\}},
\end{equation}
where $F(x,x',t)=\sqrt{G\left(B_{\sqrt{t}}(x)\right)G\left(B_{\sqrt{t}}(x')\right)}$.
\label{as:mani}
\end{assumption} 
This assumption is commonly found in the literature, such as \cite{castillo2014thomas} and \citet{10.1111/rssb.12486}, and covers a broad range of practical scenarios. For example, compact manifolds equipped with a volume form satisfy this condition.
\begin{example}
Consider the case where $\mathcal{X}$ is a compact manifold without boundary, and $G$ is the volume form $V$. In this scenario, we have $0<g(x)=1/V(\mathcal{X})<+\infty$, where $V(\mathcal{X})$ is the volume of $\mathcal{X}$. As shown in \cite{grigoryan2009heat}, the Gaussian estimate in \eqref{eq:gaussianestimate} holds for the corresponding heat kernel.
\end{example} 
 
{ For clarity in the theoretical development, we introduce the following notation for norms. For a vector $v$, the $l^2$-norm and $l^\infty$-norm are defined as $||v|| = \big(\sum_{i=1} v_i^2\big)^{1/2}$ and $||v||_\infty=\max_{i} |v_i|$, respectively. For a matrix $A$, the spectral norm and elementwise maximum norm are defined as $||A||_2=\sup_{||v||=1}||Av||$ and $||A||_{\max} = \max_{i,j} |A_{i,j}|$, respectively. For a function $f$, the $l^2(G)$-norm and $l^\infty$-norm are defined as $||f||_{2,G}=\big(\int f^2(x) dG(x)\big)^{1/2}$ and $||f||_\infty=\sup_x |f(x)|$, respectively, where $G$ is a measure.}
\subsection{One-step random walk graph Laplacian}

The first step in our theoretical framework is to evaluate the approximation error in estimating the heat kernel by using the normalized one-step random walk graph Laplacian. The concept of the one-step random walk graph Laplacian has also been explored by \citet{10.1111/rssb.12486}. However, a key distinction exists between our theoretical setup and theirs. Specifically, their work defines an integral operator associated with the continuous analogue of the graph Laplacian and employs its eigenpairs to construct the covariance function for GPs. In contrast, we directly utilize the heat kernel as the covariance function for GPs. Therefore, our prior is based on the \emph{exact} heat kernel GPs, offering a clearer framework compared to the approximate heat kernel GPs used in their prior. 

We consider a point cloud $X^n$ sampled from the manifold. Assuming that the base kernel $k$ is the squared exponential kernel with the bandwidth $\epsilon$, we construct the kernel matrix $\Bar{K}$, where each element is given by $\Bar{K}_{ij}=k(x_i,x_j)$ for $1\leq i,j\leq n$. 
The similarity matrix is then calculated as $\Bar{A}_{ij}= k(x_i,x_j)/(\Bar{K}_{i\cdot}\Bar{K}_{\cdot j})$, where $\Bar{K}_{i\cdot}$ and $\Bar{K}_{\cdot j}$ denote the row and column sums, respectively. 
The transition matrix $\Bar{Z}$ is obtained by normalizing $\Bar{A}$ so that $\Bar{Z}_{ij}=\Bar{A}_{ij}/\Bar{A}_{i\cdot}$, for $1\leq i,j\leq n$. Consequently, the one-step random walk graph Laplacian matrix $\Bar{L}$ is defined as $\Bar{L} = I - \Bar{Z}$.

Let the eigenpairs of the graph Laplacian $\Bar{L}$ be denoted by $\{(\Bar{\lambda}_{i,\epsilon},\Bar{v}_{i,\epsilon})\}_{i=1}^n$, with $||\bar{v}_{i,\epsilon}||=1$. We estimate the covariance matrix as follows:
\begin{equation*}
    \Bar{C}_{\epsilon,M,t} = n\sum_{i=1}^M\exp(-t\Bar{\lambda}_{i,\epsilon}/\epsilon^2) \Bar{v}_{i,\epsilon}\Bar{v}_{i,\epsilon}^\top.
\end{equation*}
Notably, within the exponential function, the eigenvalues $\Bar{\lambda}_{i,\epsilon}$ are scaled by $\epsilon^2$. This normalization step is essential for ensuring the correct spectral convergence of the graph Laplacian $\Bar{L}$. Further discussions on spectral convergence can be found in \citet{DUNSON2021282}. 
Let $\mathcal{L}$ denote the Dirichlet--Laplace operator on $\mathcal{X}$ with eigenpairs $\{(\lambda_i,e_i)\}_{i=1}^\infty$, where $||e_i||_{2,G}=1$. The heat kernel $p_t$ is given as $p_t(x,x')=\sum_{i=1}^\infty \exp(-t\lambda_i)e_i(x)e_i(x')$ for $x,x'\in \mathcal{X}$. Let $H_t$ be the heat kernel matrix evaluated on $X^n$.

{
\begin{assumption}
\label{as:spectrum}
The bandwidth $\epsilon$, the number of eigenpairs $M$, the point cloud size $n$ and the diffusion time $t$ satisfy the following conditions:
\begin{enumerate}
    \item Set $\Gamma_M= \min_{1\leq i\leq M} \text{dist}(\lambda_i, \{\lambda_j\}_{j\neq i})$. Suppose $\epsilon$ is small enough such that
    \begin{equation*}
        \epsilon \leq \min\left\{ \mathcal{H}_1\left(\frac{\min (\Gamma_M,1)}{\mathcal{H}_2+\lambda_M^{d/2+5}}\right)^2 , ~ \frac{1}{(\mathcal{H}_3+\lambda_M^{(5d+7)/4})^2}, ~ \frac{\mathcal{H}_4}{M^4\lambda_M^{(d-1)/2}} \right\},
    \end{equation*}
    where $\mathcal{H}_1,\mathcal{H}_2,\mathcal{H}_3$ and $\mathcal{H}_4$ are positive constants depending on $\mathcal{X}$.
    
    \item Suppose $n$ is sufficiently large so that $\left(\log n/n\right)^{1/(4d+13)} \leq \epsilon$.

    \item Suppose $\frac{\mathcal{H}_5 \log M}{M^{2/d}} \leq t$, where $\mathcal{H}_5$ is a constant depending on $\mathcal{X}$.
\end{enumerate}
\end{assumption}
This assumption, also adopted in \citet{DUNSON2021282}, imposes constraints on the relationship among $\epsilon,M,n$ and $t$. Under Assumption \ref{as:spectrum}, we have the following theorem.}

\begin{theorem}
\label{th:rwgl}
Suppose Assumptions \ref{as:mani} and \ref{as:spectrum} holds. Then, for any $t$ less than the diameter of $\mathcal{X}$, there exist constants $\mathcal{P}_1$ and $\mathcal{P}_2$ depending on the geometry of $(\mathcal{X},G)$, such that,
\begin{equation}
\label{eq:coverr}
    \|H_t-\Bar{C}_{\epsilon,M,t}\|_{\mathrm{max}} \leq \mathcal{P}_1 \epsilon + \mathcal{P}_2/M,
\end{equation} 
with probability greater than $1-n^{-2}$. 
\end{theorem}

This theorem follows directly from Theorem 3 in \citet{DUNSON2021282}, leveraging the spectral convergence of the graph Laplacian and the fast-decaying nature of the heat kernel spectrum. Equation \eqref{eq:coverr} shows that achieving accurate heat kernel estimation requires a small bandwidth $\epsilon$ and a large number of eigenpairs $M$. According to Assumption \ref{as:spectrum}, satisfying these conditions in turn necessitates a sufficiently large sample size $n$.

\subsection{Two-step random walk graph Laplacian}

{
In this subsection, we employ perturbation analysis to estimate the approximation errors associated with sparsification and subsampling. This technique is widely used in the literature, including \citet{10.1145/1557019.1557118} and \citet{10.1214/17-AOS1676}. Specifically, we define a sparse base kernel as
\begin{equation*}
    k_{\eta_{1,n}}(x,x')=k(x,x') \mathbf{1}_{\{||x-x'||<\eta_{1,n}\}},\quad x,x'\in X^n,
\end{equation*}
where $\eta_{1,n}>0$ is a threshold cutoff, and $k(x,x')=\exp\left(-||x-x'||^2/4\epsilon^2\right)$ is the squared exponential kernel. Likelihood-free kernels (\emph{e.g.}, LAE-optimized kernels) are viewed as numerical strategies to improve computational efficiency; their theoretical analysis lies beyond the scope of this work. For the kernel $k_{\eta_{1,n}}(x,x')$, we define the cross transition matrix $Z$ and the diagonal matrix $\Lambda$ as in Section \ref{sec:FGLHKE}. The corresponding two-step random walk graph Laplacian is given by $L=I-(Z\Lambda^{-1}Z^\top)^{1/2}$.
}


We begin the perturbation analysis by deriving an equivalent representation of the graph Laplacian $L$. For each data point $x_i\in X^n$, we define $u_{\tau(i)}\in I^s$ as the nearest induced point, satisfying $||u_{\tau(i)}-x_i||=\inf_{j=1,\ldots,s}||u_j-x_i||$. Let $\tilde{I}^n=\{u_{\tau(i)}\}_{i=1}^n$ represent the approximated point cloud. We define a kernel matrix $\tilde{K}$ with elements $\tilde{K}_{ij}=k_{\eta_{1,n}}(x_i,u_{\tau(j)})$ for $1\leq i,j\leq n$. Note that $\tilde{K}$ is not necessarily symmetric. Following the same procedure used for constructing the one-step random walk graph Laplacian, we derive the cross transition matrix $\tilde{Z}$ for the random walk from $X^n$ to $\tilde{I}^n$. Let $\tilde{\Lambda}$ be a diagonal matrix with diagonal elements given by $\tilde{\Lambda}_{ii}=\tilde{Z}_{\cdot i}$.

\begin{lemma}
    An equivalent representation of the two-step random walk graph Laplacian is given by $L=I - (\tilde{Z}\tilde{\Lambda}^{-1}\tilde{Z}^\top)^{1/2}$.
\end{lemma}

Consider the previously defined one-step transition matrix $\Bar{Z}$. Let $\Bar{\Lambda}$ be a diagonal matrix with entries $\Bar{\Lambda}_{ii}= \Bar{Z}_{\cdot i}$ for $1\leq i \leq n$. We define the \emph{exact} two-step random walk graph Laplacian as $I - (\Bar{Z}\Bar{\Lambda}^{-1}\Bar{Z}^\top)^{1/2}$. 

\begin{assumption}
\label{as:point}
There exists a constant $c_1\in (0,1)$ such that $\Bar{K}_{i\cdot},\Bar{K}_{\cdot i},\tilde{K}_{i\cdot},\tilde{K}_{\cdot i}\geq c_1n$ for $1\leq i\leq n$ and $n\geq 1$.
\end{assumption}

\begin{remark}
    This assumption imposes a prior constraint on the sampling distribution of the point cloud. It ensures that for each point $x_i\in X^n$, there exist sufficiently many neighboring points within a local region. Consequently, outliers are prohibited in this circumstance.
\end{remark}

{
\begin{lemma}
\label{lm:subsample}
Under Assumption \ref{as:point}, there exists a constant $C_1$ such that $ ||\Bar{Z}\Bar{\Lambda}^{-1}\Bar{Z}^\top-\tilde{Z}\tilde{\Lambda}^{-1}\tilde{Z}^\top||_2\leq C_1(\exp(-{\eta_{1,n}^2}/{4\epsilon^2}) + \eta_{2,n}/\epsilon)$,
where $\epsilon>0$ is the kernel bandwidth, $\eta_{1,n}$ is the sparsification cutoff, and $\eta_{2,n} = \sup_{1\leq i\leq n} ||x_i-u_{\tau(i)}||$ is the distortion error.
\end{lemma}
}
Lemma \ref{lm:subsample} establishes an upper bound on the approximation error of graph Laplacian in spectral norm. This result suggests that k-means clustering, which minimizes the distortion error more effectively compared to random subsampling, is likely to produce more accurate estimates in FLGP.

\begin{lemma}
\label{lm:perturbation}
Let $B_0\in S^n$ have a simple eigenvalue $\lambda_{0,B_0}$ (i.e., without multiplicity), corresponding to an eigenvector $v_0$ with $||v_0||=1$. Consider a perturbation matrix $B\in \mathbb{R}^{n\times n}$ with $||B-B_0||_2\leq \eta_0$. Then, there exists a unique eigenvalue $\lambda_B$ of $B$ associated with an eigenvector $v$ satisfying $||v||=1$, such that,
\begin{equation*}
    |\lambda_B-\lambda_{0,B_0}| \leq C_2 \eta_0,\quad ||v-v_0||\leq C_3\frac{\eta_0}{\min_{1\leq j\leq n-1}\{|\lambda_{0,B_0}-\lambda_{j,B_0}|\}},
\end{equation*}
where $C_2$ and $C_3$ are positive constants, and $\{\lambda_{j,B_0}\}_{j=1}^{n-1}$ are the eigenvalues of $B_0$ distinct from $\lambda_{0,B_0}$.
\end{lemma}

\begin{remark}
This lemma provides the perturbation results of simple eigenvalues and eigenvectors for a symmetric matrix. However, when the eigenvalues are not simple, the result generally does not hold as discussed in \citet{greenbaum2019firstorderperturbationtheoryeigenvalues}.
\end{remark}

{
\begin{lemma}
\label{lm:difftran}
Let $\gamma_n=\sup_{1\leq i,j\leq n}\Bar{A}_{i\cdot}/\Bar{A}_{j\cdot} - 1$. Then, the following bound holds: $||\Bar{Z}-\Bar{\Lambda}^{-1}\Bar{Z}^\top||_2\leq \sqrt{2}\gamma_n$.
\end{lemma}

Let $\Bar{Z}$ represent the transition matrix of the first-step random walk, and let $\Bar{\Lambda}^{-1}\Bar{Z}^\top$ represent the transition matrix of the second-step random walk. Lemma \ref{lm:difftran} implies that the discrepancy between the two transition matrices is bounded by $\gamma_n$. To provide further clarity, we present a simple scenario where $\gamma_n$ converges to zero.

\begin{example}
    For $n\geq 1$, we have $\gamma_n \leq \sup_{1\leq i,j\leq n} \Bar{K}_{i\cdot}/\Bar{K}_{j\cdot} - 1$. Specifically, when the point cloud satisfies $\Bar{K}_{i\cdot}=\Bar{K}_{j\cdot}$ for all $1\leq i,j\leq n$, it follows that $\gamma_n=0$.
\end{example}
}

{
Let $C_{\epsilon,M,t}$ be the kernel estimate based on the two-step random walk graph Laplacian under the combined effects of sparsification and subsampling. Define $\bar{\Gamma}_M$ as the minimal eigengap among the largest $M$ eigenvalues of $\Bar{Z}\Bar{\Lambda}^{-1}\Bar{Z}^\top$. Set $\eta_n = \exp(-{\eta_{1,n}^2}/{4\epsilon^2}) + \eta_{2,n}/\epsilon$.

\begin{theorem}
\label{th:tsgl}
Assume that the eigenvalues of $\Bar{Z}\Bar{\Lambda}^{-1}\Bar{Z}^\top$ are simple. 
Suppose Assumptions \ref{as:mani}, \ref{as:spectrum} and \ref{as:point} hold. For any diffusion time $t$ less than the diameter of $\mathcal{X}$, and for $\eta_n+\gamma_n \leq \min\{\bar{\Gamma}_M\sqrt{M/n},~{(M\bar{\Gamma}_M^2)}/{(n\epsilon^4)}\}$, there exist constants $\mathcal{P}_1,\mathcal{P}_2$ and $\mathcal{P}_3$ such that,
\begin{equation}
\label{eq:tsgl}
||H_t-C_{\epsilon,M,t}||_{\max} \leq \mathcal{P}_1 \epsilon + \mathcal{P}_2/M + \mathcal{P}_3M^2\sqrt{\eta_n+\gamma_n}/\epsilon^2
\end{equation}
with the probability greater than $1-n^{-2}$.
\end{theorem}

\begin{remark}
The error bound in Equation \eqref{eq:tsgl} consists of two components. The term $\mathcal{P}_1 \epsilon + \mathcal{P}_2/M$ accounts for the approximation error of the one-step random walk graph Laplacian. The term $\mathcal{P}_3M^2\sqrt{\eta_n+\gamma_n}/\epsilon^2$ corresponds to the additional errors introduced by sparsification and subsampling, which vanishes as $\eta_n$ and $\gamma_n$ decrease to zero.
\end{remark}

This theorem follows directly from Theorem \ref{th:rwgl} and Lemmas \ref{lm:subsample}, \ref{lm:perturbation}, and \ref{lm:difftran}. It provides an upper bound on the approximation error in estimating the heat kernel using the two-step random walk graph Laplacian. Note that Theorem \ref{th:tsgl} does not establish that the sparsification technique improves the convergence rate. However, simulation studies show that graph-based methods employing a sparse base kernel achieve better regression performance. A thorough analysis of this phenomenon is left for future work.
}

\subsection{Predictive errors in Gaussian processes}
For a given test point $x^*$, we denote $f(x^*)$ as $f^*$. 
Consider two covariance matrices $\Sigma^1$ and $\Sigma^2$ for the Gaussian process $f$ evaluated at the points $\{x_1,\ldots,x_m,x^*\}$. For $i=1,2$, we partition $\Sigma^i$ in blocks as $\begin{pmatrix}
    \Sigma^i_{f^mf^m}&\Sigma^i_{f^mf^*}\\ \Sigma^i_{f^*f^m} & \Sigma^i_{f^*f^*}\\
\end{pmatrix}$, where $\Sigma^i_{f^mf^m}\in \mathbb{R}^{m\times m}$  is the covariance matrix at the training points. The conditional distribution is given by $\pi_{\Sigma^i}(f^*|f^m)\sim N(\mu_i,\sigma_i^2)$ with 
\begin{equation*}
    \mu_i = \Sigma^i_{f^*f^m}(\Sigma^i_{f^mf^m})^{-1}f^m,\quad \sigma_i^2 = \Sigma^i_{f^*f^*} - \Sigma^i_{f^*f^m}(\Sigma^i_{f^mf^m})^{-1}\Sigma^i_{f^mf^*}.
\end{equation*}

\begin{lemma}
\label{lm:cnorm}
Assume that $\|\Sigma^1-\Sigma^2\|_{\mathrm{max}}\leq \delta < 1$. Then there exist constants $c_2,c_3>0$ such that, for $\mu_1,\mu_2$ and $\sigma_1^2,\sigma_2^2$,
\begin{equation*}
    \begin{split}
        |\mu_1-\mu_2|\leq \frac{\delta m||f^m||_\infty\left\{(c_2+3\lambda)|| \Sigma^1_{f^*f^m} ||_\infty + \lambda \right\}}{\lambda(\lambda+c_2\delta)},\\
        |\sigma_1^2 - \sigma_2^2| \leq \delta + \frac{2\delta m||\Sigma^1_{f^*f^m}||_\infty\left\{(c_3+3\lambda)||\Sigma^1_{f^*f^m}||_\infty+\lambda\right\}}{\lambda(\lambda+c_3\delta)},
    \end{split}
\end{equation*}
where $\lambda$ is the smallest eigenvalue of $\Sigma^1_{f^mf^m}$.
\end{lemma}

\begin{lemma}
\label{lm:postnorm}
Suppose $|\mu_1-\mu_2|\leq \delta_1$ and $|\sigma_1^2-\sigma_2^2|\leq \delta_2$. Then we have
\begin{equation*}
    |\mathbb{E}_1\{l(f^*)|f^m\} - \mathbb{E}_2\{l(f^*)|f^m\}| \leq ||l'||_\infty( \delta_1 + \mathbb{E}|\xi|\sqrt{\delta_2} ),
\end{equation*}
where $\xi\sim N(0,1)$.
\end{lemma}


\begin{theorem}
\label{th:pred}
Assume $\|H_t-C_{\epsilon,M,t}\|_{\mathrm{max}}\leq \delta < 1$. Then there exists $\mathcal{P}_4>0$ such that,
\begin{equation*}
    \max_{m+1\leq i\leq n}\left|\mathbb{E}\{\varphi(x_i)|f^m,H_t\} - \mathbb{E}\{\varphi(x_i)|f^m,C_{\epsilon,M,t}\}\right| \leq \mathcal{P}_4 ||l'||_\infty\sqrt{\delta},
\end{equation*}
where $\mathcal{P}_4$ depends on $m,~f^m$ and $H_t$.
\end{theorem}

\begin{remark}
    In the normal regression model, where the GP is the conjugate prior and $\varphi=f$, Lemma \ref{lm:cnorm} implies that $\max_{m+1\leq i\leq n}\left|\mathbb{E}\{\varphi(x_i)|Y^m,H_t\} - \mathbb{E}\{\varphi(x_i)|Y^m,C_{\epsilon,M,t}\}\right|$ is bounded by the same bound as $|\mu_1-\mu_2|$, with the substitution of $||f^m||_\infty$ by $||Y^m||_\infty$. However, for other non-conjugate models, the situation is probably very complicated, and its thorough exploration is beyond the scope of this article.
\end{remark}

\subsection{Posterior contraction rates in the canonical exponential family}


{ The convergence of heat kernel Gaussian processes on manifolds was previously studied by \citet{castillo2014thomas}. However, their analysis was restricted to relatively simple likelihoods, such as Gaussian process regression. In this work, we extend the existing results on posterior contraction rates to a more general setting of canonical exponential family likelihoods, a scenario that, to the best of our knowledge, has not yet been addressed.} 
Additionally, since multi-parameter CEF models can often be expressed as compositions of one-parameter CEF models, we restrict our discussion to the one-parameter case for simplicity.


Let $\{(x_i,y_i)\}_{i=1}^m$ be independent and identically distributed (\emph{i.i.d.}) observations from $(X,Y)\in \mathcal{X}\times \mathcal{Y}$, where $\mathcal{X}$ is a $d$-dimensional submanifold of $\mathbb{R}^p$ and $\mathcal{Y}\subset \mathbb{R}$. We consider the product measure $\nu=V\times\mu$ on $\mathcal{X}\times \mathcal{Y}$, with $\mu$ representing either the Lebesgue measure or the counting measure. Assume that $X\sim G$ and $\pi_\theta(y|x)\sim \text{CEF}(\theta(x))$. 
Consequently, for $x\in \mathcal{X}$, the relationships $J'(\theta(x))=\varphi(x)=l\circ f(x)$ and $J''(\theta(x)) = \text{Var}_{\theta(x)}(\kappa(Y))$ hold. 
If $J'$ is invertible, defining $\zeta=(J')^{-1}\circ l$, we have $\theta=\zeta\circ f$.

\begin{assumption}
Suppose that the mapping $J:\mathbb{R}\to \mathbb{R}$ belongs to the space $C^2(\mathbb{R})$. Consider the minimal exponential family such that $J'$ is invertible. Let $\varphi=l\circ f$ and $\theta = \zeta \circ f$. Suppose that the functions $l$ and $\zeta$ are continuously differentiable, and their derivatives $l'$ and $\zeta'$ are uniformly bounded.
\label{as:link}
\end{assumption}

Assumption \ref{as:link} holds for a wide range of models, including Gaussian process regression and classification. 
Applying the chain rule to $\zeta=(J')^{-1}\circ l$, we obtain $\zeta'(f(x))=l'(f(x))/J''(\theta(x))$ for $x\in \mathcal{X}$. If $J''$ has a positive lower bound, this assumption is equivalent to controlling the growth rate of $l$. If the link function $l$ fluctuates too rapidly, even an accurate estimation of the latent function $f$ may lead to large predictive errors for $\varphi$, as highlighted in Theorem \ref{th:pred}. Therefore, it is nearly impossible to derive posterior contraction rates without imposing some conditions on $l'$. 

To establish the contraction rate, we begin by deriving relationships among several statistical metrics within the one-parameter CEF. 
Consider two latent functions $f_1,f_2$, and let $\pi_{f_1}$ and $\pi_{f_2}$ represent the joint density functions of $(X,Y)$ under the dominant measure $\nu$. The Kullback-Leibler divergence between these densities is defined as $\text{KL}(\pi_{f_1};\pi_{f_2})=\mathbb{E}_{\pi_{f_1}}\log(\pi_{f_1} / \pi_{f_2})$. Additionally, the $i$-th Kullback-Leibler variation is defined as $\text{V}_{i,0}(\pi_{f_1};\pi_{f_2})=\mathbb{E}_{\pi_{f_1}}\left|\log(\pi_{f_1} / \pi_{f_2})-\text{KL}(\pi_{f_1};\pi_{f_2})\right|^i$. The Hellinger distance between $\pi_{f_1}$ and $\pi_{f_2}$ is defined as $d_H(\pi_{f_1},\pi_{f_2})=\left\{\mathbb{E}_\nu(\sqrt{\pi_{f_1}}-\sqrt{\pi_{f_2}})^2\right\}^{1/2}$. 

\begin{lemma}
Suppose Assumption \ref{as:link} holds. If $\pi_{f_1}(y|x)$ and $\pi_{f_2}(y|x)$ belong to one-parameter CEF, then the following results hold:
\begin{enumerate}
    \item For given functions $f_1$ and $f_2$, suppose that they take values in a bounded set $\Omega \subset \mathbb{R}$. There exist positive constants $C_4(\Omega)$ and $C_5(\Omega)$, such that
    \begin{equation*}
        \text{KL}(\pi_{f_1};\pi_{f_2}) \leq C_4(\Omega)||f_1-f_2||_{\infty}^2,\qquad \text{V}_{2,0}(\pi_{f_1};\pi_{f_2}) \leq C_5(\Omega) ||f_1-f_2||_{\infty}^2.
    \end{equation*}
    \item There exists a uniform constant $C_6$, such that
    \begin{equation*}
        d_H^2(\pi_{f_1},\pi_{f_2}) \leq C_6||f_1-f_2||_{\infty}^2.
    \end{equation*}
\end{enumerate}
\label{lm:metric}
\end{lemma}
The proof employs Taylor expansion techniques to derive the upper bounds for the KL divergence, KL variation, and Hellinger distance in terms of the $l^\infty$-norm difference between the functions $f_1$ and $f_2$. The assumption of the boundedness of $\Omega$ is naturally satisfied when $f_1$ and $f_2$ are continuous functions defined on a compact set. Lemma \ref{lm:metric} plays a pivotal role in deriving posterior contraction rates for one-parameter CEF models.

The diffusion time $t$ of the heat kernel plays a role analogous to the bandwidth parameter in the squared exponential kernel. If the time $t$ is fixed, the sample paths of the corresponding GP are almost surely infinitely smooth, potentially leading to slow posterior contraction rates. To achieve minimax contraction rates for the posterior, we introduce a prior on $t$ that allows adaptation to the unknown function's regularity. Accordingly, we consider an adaptive heat kernel Gaussian process $\{f(x),x\in \mathcal{X}\}$ defined as follows:
\begin{equation}
    f\mid T \sim \mathcal{GP}(0,p_T(x,x')),
\end{equation}
where $T$ is a random diffusion time with density $\pi(t)\propto t^{-a_0}\exp\{-t^{-d/2}\log^{1+d/2}(1/t)\}$ for $a_0>1$.  This choice of  prior is widely used in the literature, including the seminal work of \citet{castillo2014thomas}. 
We now state the main result regarding the contraction rates of adaptive heat kernel GPs on manifolds involving CEF likelihoods.

\begin{theorem}
Suppose the observations $\{(x_i,y_i)\}_{i=1}^m$ are independently and identically distributed from the population $\pi_f(x,y)=\pi_f(y|x)g(x)$, where $f$ is the latent function. Let the prior on $f$ be the heat kernel GP with an adaptive diffusion time. Suppose the marginal density $g(x)$ satisfies Assumption \ref{as:mani}, and the conditional density $\pi_f(y|x)$ belongs to one-parameter CEF models satisfying Assumption \ref{as:link}. Then, for a true function $f_0$ in the Besov space $B^\beta_{\infty,\infty}(\mathcal{X})$ with $\beta>0$, the posterior contraction rate at $f_0$ \emph{w.r.t.} $d_H$ is $\epsilon_m\sim (\log m/m)^{2\beta/(2\beta+d)}$. That is, as $m\to \infty$, for each $M_m\to \infty$,
\begin{equation}
    \mathbb{E}_{f_0}\Pi\left\{f,~d_H(\pi_f,\pi_{f_0})\geq M_m\epsilon_m\mid X^m,Y^m\right\} \to 0,
\end{equation}
where $\Pi\{~\cdot\mid X^m,Y^m\}$ denotes the posterior distribution.
\label{th:cont}
\end{theorem}
This theorem follows from the relationships among statistical distances established in Lemma \ref{lm:metric}, combined with the property of heat kernel GPs as discussed in \citet{castillo2014thomas}. The true function space $B^\beta_{\infty,\infty}(\mathcal{X})$ is a Besov space with smoothness parameter $\beta$; see Section 9.1 of \citet{castillo2014thomas} for a formal definition. 

{The heat kernel GPs on manifolds achieve an optimal posterior contraction rate, up to a logarithmic term in the minimax sense, that depends on the intrinsic dimension of the manifold. However, we note that this optimal rate alone cannot distinguish the data efficiency between our approach and extrinsic GPs. Specifically, for input data lying on a submanifold embedded in a Euclidean space, \citet{10.1214/15-AOS1390} demonstrated that GPs with ambient squared exponential kernels surprisingly attain the same asymptotic contraction rate. Nonetheless, as highlighted in \citet{rosa2023posterior} and confirmed by our empirical analysis, intrinsic GPs can significantly outperform extrinsic GPs on highly non-Euclidean domains due to differences in constant factors. }


In Theorem \ref{th:cont}, the diffusion time $t$ is treated as a random variable to facilitate adaptation. However, in practical implementations of FLGP, $t$ is optimized by maximizing the marginal likelihood, rendering the approach an empirical Bayesian method. The gap between the theoretical framework and practical methodology presents an interesting avenue for future research. 
\section{Experiments}

In this section, we conduct a comparative evaluation of the performance across various kernel-based methods. Our analysis involves four distinct versions of the FLGP method, each defined by a different combination of base kernels and subsampling strategies. These versions include: the squared exponential kernel with random subsampling (SRFLGP), the squared exponential kernel with k-means subsampling (SKFLGP), local anchor embedding with random subsampling (LRFLGP), and local anchor embedding with k-means subsampling (LKFLGP). { Additionally, we consider a special case of FLGP without subsampling as an ablation study, where the induced point set $I^s$ is the point cloud $X^n$ and the squared exponential kernel is the base kernel; we denote it as FLGP-X. }

Alongside FLGP, our analysis includes the Euclidean Gaussian Processes (EGP) with squared exponential kernels, GLGP, 
and a Nyström approximation of GLGP (denoted GLGP$^*$). The Nystr\"om approximation is a widely recognized technique to enhance the scalability of GPs, as discussed by \citet{quinonero2005unifying}. In Section D of supplementary materials, we detail our implementation of the Nystr\"om approximation based on the work of \citet{10.1111/rssb.12486}. 

For the spectral decomposition of graph Laplacian, we utilize a Lanczos-type algorithm \citep{doi:10.1137/04060593X}. Specifically, the \texttt{RSpectra} package---an \texttt{R} interface to the \texttt{C++} library \texttt{Spectra}---is employed as our solvers for large-scale eigendecomposition and singular value decomposition tasks. 

We present three examples to evaluate the performance. The first example involves a synthetic dataset, where we simulate multiple concentric circles in $\mathbb{R}^2$. The second example utilizes the MNIST dataset, which consists of handwritten-digit images. The third example pertains to the reconstruction of fMRI images from the Human Connectome Project (HCP). In the MNIST and HCP tasks, the domain manifolds are unknown and potentially intricate. Computationally, the first two examples are executed on a machine equipped with 80GB of RAM and a 3.70GHz AMD Ryzen 5 5600X 6-core processor. The HCP application is run in parallel on the Great Lakes computing cluster at the University of Michigan, using one CPU core from a 3.0GHz Intel Xeon Gold 6154 processor and 10GB of RAM per subject.

Moreover, GPs are valued for their capacity to deliver well-calibrated uncertainty estimates. To evaluate the quality of uncertainty quantification in our experiments, we compute the negative log-likelihood (NLL) of the test data as a calibration metric.

In Section E of supplementary materials, we conduct a sensitivity analysis of the predefined hyperparameters $s$, $r$ and $M$ in FLGP.  
In Section F of supplementary materials, we design a further experiment to provide insights into how different sampling distributions of the point cloud may affect the model performance. In Section G of supplementary materials, we implement a fast regression algorithm for the proposed method when the number of labeled observations is large.

\subsection{Simulation on synthetic concentric circles}

\begin{figure}
    \centering
    \includegraphics[width=0.4\textwidth]{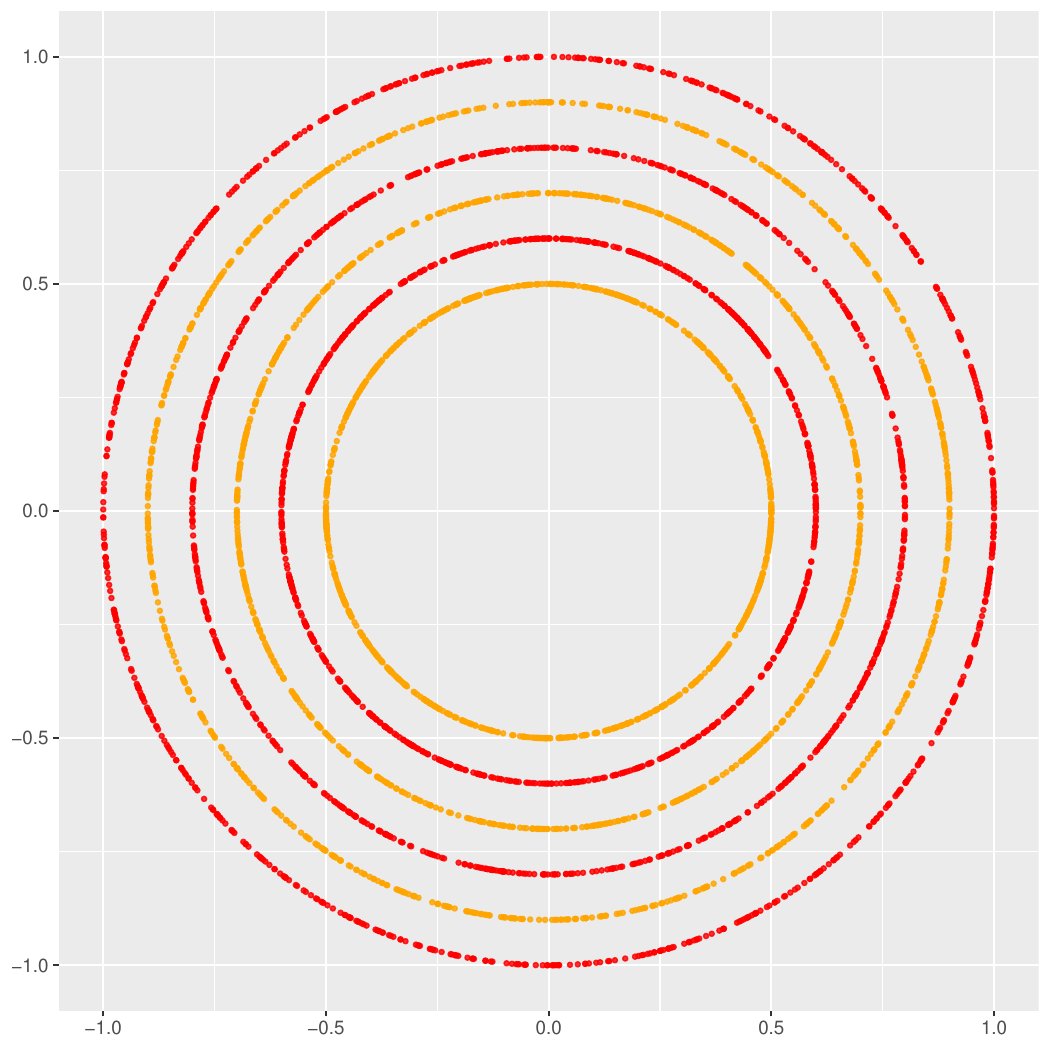}
    \caption{The concentric circles in $\mathbb{R}^2$, where red points are of Class 0 and orange points are of Class 1.}
    \label{fig:circle}
\end{figure}

In this example, we construct a dataset based on six concentric circles in $\mathbb{R}^2$. The circles are denoted sequentially from the innermost to the outermost as $S_1,S_2,S_3,S_4,S_5,S_6$, with the collective set $S=\bigcup_{i=1}^6 S_i$ representing a one-dimensional submanifold embedded in $\mathbb{R}^2$. For data generation, we begin by uniformly sampling $n/6$ points uniformly on each of the six circles, resulting in a total of $n$ points that form the point cloud $X^n$. Subsequently, we randomly sample $m$ points from $X^n$ to serve as the labeled dataset. The points located on $S_1,S_3,S_5$ are assigned to Class 1, while those on on $S_2,S_4,S_6$ are assigned to Class 0. A visual representation of this synthetic dataset, consisting of concentric circles, is provided in Figure \ref{fig:circle}.

We explore three scenarios for data generation by varying $n$ in the range $\{3000,9000,900000\}$ while keeping $m=50$. The number of eigenpairs used is set to $M=100$. For FLGP with subsampling, the hyperparameters are configured as $r=3$ and $s=600$. For FLGP-X, the hyperparameter $r$ is set to 20 when $n=3000$, and $60$ when $n=9000$. Due to substantial computational requirements, GLGP and FLGP-X 
are not applied in the scenario where $n=900000$. Each scenario is repeated $20$ times.

The performance of various GP models, including error rates and negative log-likelihoods (NLL) on unlabeled data, is summarized in Table \ref{tab:circleperf}. Generally, kernel approaches utilizing intrinsic geometry outperform those based on extrinsic kernels. Intrinsic GP methods exhibit decreasing error rates and lower NLL values as the point cloud size $n$ increases, highlighting the importance of sufficient data for reliable heat kernel estimation. { Notably, the FLGP-X method achieves the best performance, outperforming even the standard GLGP. This result suggests that sparsity improves accuracy by promoting a local connectivity pattern in the graph Laplacian, which more effectively captures the intrinsic geometry.}
Our proposed FLGP, particularly with k-means subsampling, attains lower error rates and NLLs than the Nystr\"om approximation of GLGP (GLGP$^*$). Moreover, comparisons among different subsampling strategies within FLGP underscores the advantage of k-means clustering in identifying induced points that better reflect the domain's geometry over random subsampling. 
{ Finally, FLGP is less sensitive to the choice of the bandwidth parameter $\epsilon$ compared to standard GLGP, which requires an extremely small $\epsilon$ to achieve reasonable accuracy for this task---at the cost of numerical instability in the eigendecomposition of the graph Laplacian.}

\begin{table}
\caption{ Error rates (\%) and negative log likelihoods (NLL) in the concentric circles with $m=50$ labeled observations, $n=3000,9000,900000$ points (labeled + unlabeled).}
\label{tab:circleperf}
\centering
\small
{
\begin{tabular}{@{}ccccccc@{}}
\toprule
                  & \multicolumn{3}{c}{Error Rates}   & \multicolumn{3}{c}{NLL}              \\ \midrule
$n$                 & 3000      & 9000      & 900000    & 3000       & 9000       & 900000     \\ \midrule
EGP               & 45.7(1.6) & 46.7(2.5) & 45.9(2.0)   & 0.67(0.01) & 0.68(0.01) & 0.67(0.01) \\
GLGP              & 5.0(3.6)    & 3.8(3.4)  &    -       &  0.37(0.01) & 0.34(0.01) &     -       \\
GLGP$^*$ & 22.2(3.2) & 14.9(4.4) & 10.4(5.1) & 0.48(0.01) & 0.43(0.02) & 0.40(0.02)  \\
FLGP-X & {\bf \textless{}0.1(\textless{}0.1)} & {\bf \textless{}0.1(\textless{}0.1)} & - & {\bf 0.19(\textless{}0.01)} & {\bf 0.18(0.01)} & - \\
SRFLGP            & 16.9(6.3) & 9.1(4.5)  & 10.7(5.5) & 0.47(0.03) & 0.40(0.02)  & 0.41(0.03) \\
SKFLGP            & { 1.0(2.6)}    & { \textless{}0.1(\textless{}0.1)}      & {\bf 0.1(0.4)}  & { 0.21(0.03)} & { 0.19(0.01)}    & {\bf 0.19(0.01)} \\
LRFLGP            & 29.1(4.6) & 28.7(4.6) & 28.3(3.6) & 0.58(0.02) & 0.57(0.03) & 0.57(0.02) \\
LKFLGP            & 5.8(3.3)  & 2.4(3.1)  & 2.1(3.0)    & 0.35(0.04) & 0.29(0.04) & 0.26(0.04) \\ \botrule
\end{tabular}
}
\begin{tablenotes}
    \item  {\bf Note:} EGP is the Euclidean Gaussian Process with squared exponential kernels; GLGP is the Graph Laplacian Gaussian Process; GLGP$^*$ is the Nystr\"om approximation of GLGP; FLGP-X is the special case of FLGP without subsampling; SRFLGP, SKFLGP, LRFLGP and LKFLGP are four versions of FLGP with different base kernels and subsampling strategies.
\end{tablenotes}
\end{table}

Regarding efficiency, Table \ref{tab:time} reports the computational times for the different GPs. The results indicate that the eigendecomposition of the graph Laplacian---despite using an iteration-based solver---remains computationally intensive, rendering GLGP impractical for large-scale datasets. As the point cloud size $n$ increases, our FLGP method demonstrates significantly faster speed than other manifold learning techniques. 
For instance, at $n=900000$, LKFLGP exhibits more than a 10-fold reduction in computation time compared to GLGP$^*$. 

\begin{table}
\caption{Computational time (seconds) in concentric circles and MNIST.}
\label{tab:time}
\centering
\small
{
\begin{tabular}{@{}ccccccc@{}}
\toprule
Dataset           & \multicolumn{3}{c}{Circles} & \multicolumn{3}{c}{MNIST} \\ \midrule
$n$                 & 3000   & 9000     & 900000  & 7000   & 70000  & 700000  \\ 
$m$                 & 50     & 50       & 50      & 200    & 200    & 200     \\ \midrule
EGP               & {\bf 0.12}   & 0.37     & 35.17   & 4.7    & 40.6   & 401.7   \\
GLGP              & 7.49   & 102.68   &     -    & -     & -     & -     \\
GLGP$^*$ & 3.18   & 6.89     & 589.14  & 32.9   & 171.3  & 1466.1  \\
FLGP-X & 6.65 & 20.04 & - & - & - & - \\
SRFLGP            & 2.67   & 2.72     & 117.19  & 22.8   & 35.1   & 193.4   \\
SKFLGP            & 6.62   & 3.53     & 123.06  & 27.5   & 51.5   & 230.7   \\
LRFLGP            & 0.14   & {\bf 0.23}     & {\bf 17.45}   & {\bf 4.2}    & {\bf 12.6}   & {\bf 100.7}   \\
LKFLGP            & 0.15   & 0.26     & 19.77   & 5.0      & 27.9   & 154.9   \\ \botrule
\end{tabular}
}
\end{table}

\begin{figure}
    \centering
    \includegraphics[width=0.8\textwidth]{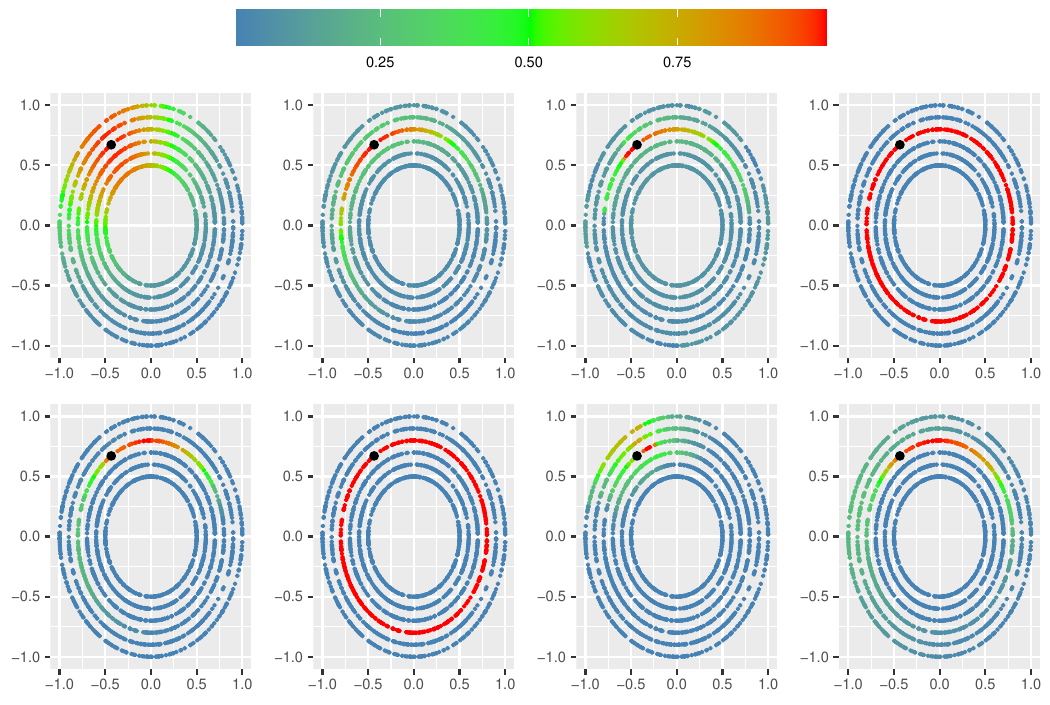}
    \caption{ The prior covariance between the point cloud and a reference point with $m=50$ and $n=9000$ in the circle example, where the black point is the reference point. Top panels: EGP, GLGP, GLGP$^*$, FLGP-X; Bottom panels: SRFLGP, SKFLGP, LRFLGP, LKFLGP.}
    \label{fig:cov_circles}
\end{figure}

To delve into the underlying mechanisms, we present the prior covariance distributions of various kernels in Figure \ref{fig:cov_circles}. For enhanced visualization, these covariance values are normalized to $[0,1]$. Given the highly non-Euclidean nature of the domain geometry, the squared exponential kernel is observed to be inadequate in capturing the true similarity between points, as its activation strength depends on Euclidean distance. In contrast, the heat kernel estimation effectively encapsulates the intrinsic structure, demonstrating that only points within the same circle exhibit high prior covariance with respect to the reference point. These distinct covariance distributions are instrumental in contributing to the performance disparities among the different approaches.

\subsection{Application to handwritten digit data MNIST}

The MNIST database comprises 60,000 training and 10,000 test image-label pairs, with each image being a $28 \times 28$ pixel matrix representing a handwritten digit from 0 to 9.  
To assess performance on a larger scale, we create an augmented MNIST dataset by randomly cropping the original images within a scale range of $[0.8,1.0]$ for the crop area, followed by resizing them to the original dimensions. After performing this augmentation 9 times for each image, we expand the dataset to a total of 700,000 samples, including both original and augmented images. 

We explore three scenarios for data generation, varying the number of observations $n$ in the range ${7000, 70000, 700000}$, while keeping the number of labeled points $m$ fixed at 200. To reduce the domain dimension, we apply Principal Component Analysis (PCA) to the grayscale images, retaining the first 100 principal components as predictors. 
As the standard GLGP and FLGP-X are time-consuming in this large-scale data set, we include only GLGP$^*$ as a benchmark in this experiment. For our FLGP, the hyperparameters are set to $M=100$, $r=3$, and $s=1000$. Each scenario is repeated 10 times.

Table \ref{tab:mnistperf} presents the error rates and negative log-likelihoods (NLL) of various GPs in the MNIST example. Consistent with the concentric circle case, manifold learning approaches outperform Euclidean-based approaches across all scenarios, underscoring the importance of leveraging intrinsic geometry for predictive performance. 
Our FLGP achieves substantially lower error rates and NLLs compared to the Nystr\"om approximation (GLGP$^*$), suggesting that FLGP constructs a better reduced-rank approximation of the graph Laplacian. Table \ref{tab:mnistperf} also indicates that cluster centers offer a notable advantage over random induced points. Additionally, FLGP with an LAE-like kernel performs comparably to FLGP with squared exponential kernels, while requiring less processing time, highlighting the utility of constructing likelihood-independent base kernels.

\begin{table}
\caption{Error rates (\%) and negative log likelihoods (NLL) in MNIST with $m=200$ labeled observations, $n=7000,70000,700000$ points (labeled + unlabeled).}
\label{tab:mnistperf}
\centering
\small
\begin{tabular}{@{}ccccccc@{}}
\toprule
                  & \multicolumn{3}{c}{Error Rates} & \multicolumn{3}{c}{NLL}           \\
                  \midrule
$n$                & 7000      & 70000     & 700000$^{1}$    & 7000      & 70000     & 700000    \\ \midrule
EGP               & 26.4(2.5) & 27.4(3.3) & 33.9(3.5) & 2.5(0.02) & 2.5(0.02) & 2.6(0.02) \\
GLGP$^*$ & 23.0(3.0)     & 12.8(1.9) & 31.7(1.9) & 2.2(0.06) & 1.7(0.03) & 2.3(0.04) \\
SRFLGP            & 14.5(2.6) & 13.6(2.1) & 20.4(1.6) & 1.7(0.03) & 1.7(0.03) & 1.9(0.03) \\
SKFLGP            & {\bf 9.5(1.3)}  & 7.1(2.1)  & {\bf 12.6(1.6)} & {\bf 1.6(0.02)} & {\bf 1.5(0.02)} & 1.6(0.03) \\
LRFLGP            & 15.1(1.1) & 15.2(1.8) & 22.2(1.7) & 1.7(0.03) & 1.7(0.03) & 1.9(0.03) \\
LKFLGP            & 9.7(1.0)    & {\bf 7.0(1.4)}    & 13.1(1.9) & 1.6(0.03) & 1.5(0.03) & {\bf 1.6(0.02)} \\ \botrule
\end{tabular}
\begin{tablenotes}
\item[$^{1}$] In the augmented MNIST, the clustering method is mini-batch k-means \citep{sculley2010web}.
\end{tablenotes}
\end{table}

\subsection{Application to fMRI data from HCP}\label{sec:hcp}

We further illustrate our method with analysis of task-evoked functional Magnetic Resonance Imaging (fMRI) data in the Human Connectome Projects (HCP-1200 release). The task-evoked fMRI is a technique that indirectly measures brain activities by monitoring oxygen-level changes in the blood flow when the subject performs specific tasks. In this study, we focus on the $2$-back versus $0$-back (2bk-0bk) contrast maps during the working memory $N$-back tasks for $917$ subjects in the HCP, where all the images have been registered on the standard MNI 2mm brain template of resolution $91\times 109 \times 91$. We refer the reader to \citet{van2013wu} for a detailed description of the contrast maps and the $N$-back tasks. Briefly speaking, the contrast map for a single subject is a $3$-dimensional tensor that reflects the difference of the fMRI signals at each voxel under different task conditions in the brain of the subject.

\subsubsection{Voxel-by-voxel image reconstruction}\label{sec:hcp_reconstruction}

{ Incomplete intensity values are a common issue in real-world neuroimaging, often arising from limited image acquisition or susceptibility artifacts \citep{mulugeta2017methods}. Imputation for missing data in fMRI studies remains an open problem, and existing methods cannot adequately address this challenge \citep{lu2025new}.} 

{ In this section, we aim to reconstruct the 2bk-0bk contrast map on a voxel-by-voxel basis using Gaussian process regression, a framework that naturally aligns with the imputation of missing data.} The voxel-level contrast intensity serves as the response variable, and the three-dimensional voxel location acts as the predictor. 
Since only specific brain neurons are activated during a given task, we define an activation region rather than considering the entire brain as the regression domain. Moreover, due to inter-subject heterogeneity, the activation region is determined individually for each subject.

\begin{figure}
    \centering
    \includegraphics[width=0.8\textwidth]{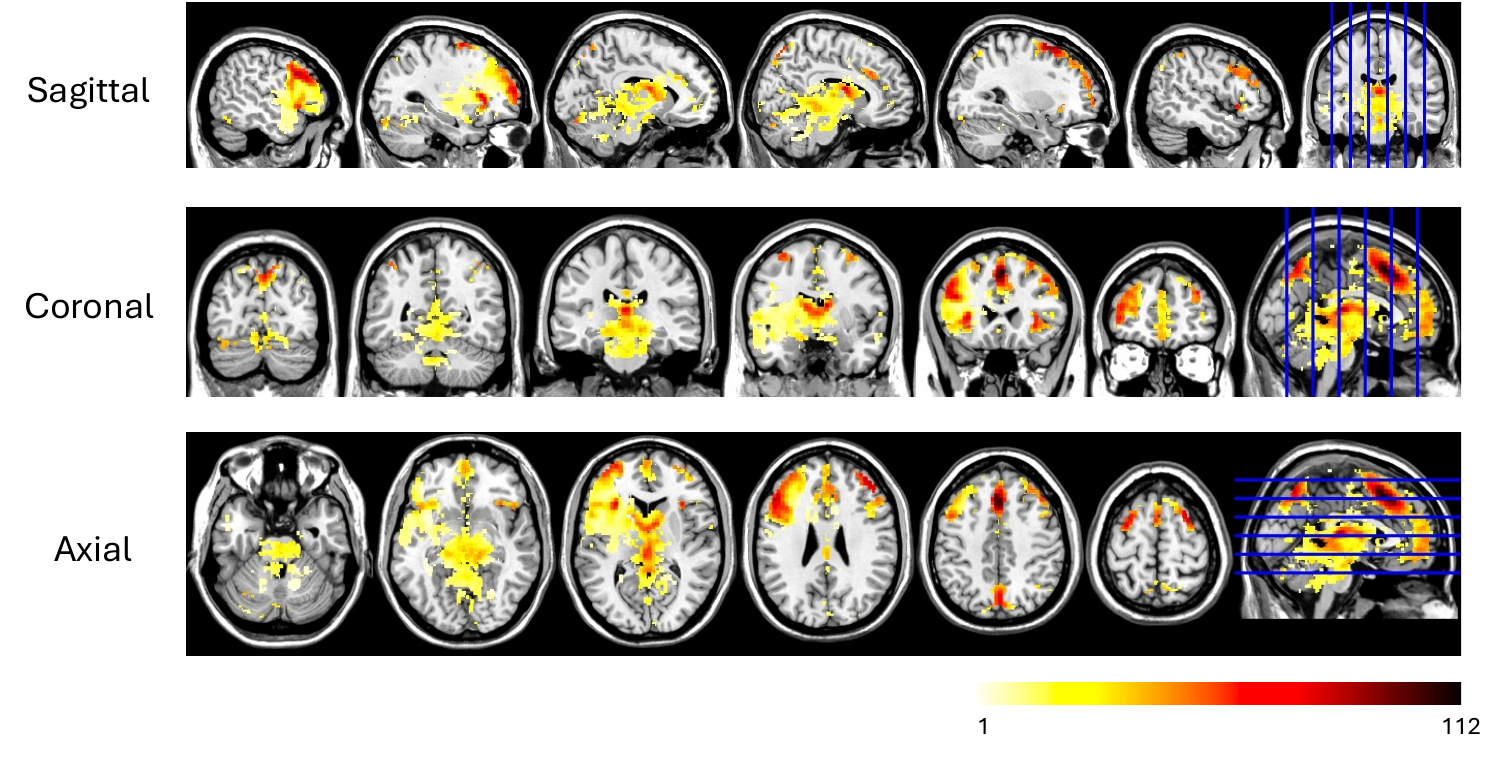}
    \caption{Activation count map for the 2bk-0bk contrast, shown in sagittal, coronal, and axial views. The count value at each voxel represents the number of subjects where the voxel is activated.}
    \label{fig:activation_map}
\end{figure}

To remove outlier voxels, our analysis is restricted to locations where at least two subjects exhibit absolute intensities exceeding the $95\%$ standard Gaussian quantile, referred to as effective voxels. For each individual, the subject-specific activation region comprises effective voxels with absolute intensity values surpassing the $75\%$ standard Gaussian quantile. The substantial variability across individuals is visually demonstrated in the activation count map (Figure \ref{fig:activation_map}), which shows that many voxels are activated in fewer than a quarter of the total subjects while remaining inactive in the others.

In this example, we compare the performance of EGP, GLGP$^*$, and FLGP, using LKFLGP as a representative of the FLGP family. LKFLGP is selected because it offers the best trade-off between performance and efficiency among the FLGP variants, as shown by simulation studies. The standard GLGP method is excluded due to its prohibitive computational cost. Given the weak signal characteristics of fMRI data, statistical models are prone to severe overfitting, leading to unreliable predictions when the number of labeled observations, $m$, is small. To address this challenge, we consider 134 subjects with at least 5000 activated voxels and evaluate three scenarios with labeled ratios of $m/n=0.2,0.3$ and $0.4$. The hyperparameters of LKFLGP are set to $M=500$, $r=3$, and $s=1500$. Each scenario is repeated 10 times for each subject.

The experimental results, summarized in Table \ref{tab:hcp}, present the average voxel-wise mean squared error (MSE), negative log-likelihood (NLL) and computational time (Time) across subjects and repetitions. To mitigate the influence of outliers, predictive errors within the top or bottom $1\%$ are excluded from the analysis. In all scenarios, FLGP achieves the lowest MSE and NLL, while EGP exhibits the highest. This observation indicates that incorporating intrinsic geometry enhances performance, with FLGP offering a superior estimation of the brain manifold compared to GLGP$^*$. Furthermore, due to its low-rank structure and sparsity, FLGP emerges as the most computationally efficient approach, requiring significantly less processing time than the other methods. 

\begin{table}
\caption{{ Mean square errors (MSE) ($\times 10^{-1}$), negative log likelihood (NLL) ($\times 10^{-1}$) and computational time (seconds) in the HCP image reconstruction task.}}
\label{tab:hcp}
\centering
\resizebox{\textwidth}{!}{%
\begin{tabular}{cccccccccc}
\toprule
Labeled Ratio          & \multicolumn{3}{c}{0.2}   & \multicolumn{3}{c}{0.3} & \multicolumn{3}{c}{0.4} \\
\midrule
    Method              & MSE   & NLL   & Time  & MSE   & NLL   & Time  & MSE   & NLL & Time    \\
\midrule
EGP                     & 1.43(1.53)  & 3.16(2.50)    & 4.2     & 1.35(1.45)  & 2.86(3.48)      & 14.0      & 1.30(1.41)  & 2.69(3.47)    & 32.3     \\
GLGP$^*$                & 1.04(0.81)  & 2.00(2.77)  & 75.5    & 0.97(0.77)    & 1.64(2.80)      & 80.8      & 0.94(0.76)  & 1.47(2.83)     & 85.7     \\
FLGP                    & {\bf 0.83(0.43)}    & {\bf 1.34(2.06)}    & {\bf 2.9} & {\bf 0.75(0.39)}    & {\bf 0.87(2.07)}    & {\bf 3.5} & {\bf 0.73(0.38)}    & {\bf 0.69(2.07)}    & {\bf 3.9}     \\
\botrule
\end{tabular}%
}
\end{table}

\subsubsection{Prior correlation heterogeneity across subjects}

In this subsection, we examine the heterogeneity in prior correlations estimated by FLGP across subjects and emphasize its ability to adapt to individual-specific brain coactivation. Conventional fMRI analyses typically rely on Euclidean geometry, which fails to capture the intrinsic structure of the human brain. Although cortical surface models \citep{fischl1999cortical} allow for analyzing brains in a lower-dimensional space, these surfaces are pre-defined and constructed uniformly for all brains, limiting their adaptability to individual subjects or specific tasks. In contrast, FLGP can estimate a subject-specific manifold, offering a flexible framework to explore inter-individual differences.

\begin{figure}
    \centering
    \includegraphics[width=0.8\textwidth]{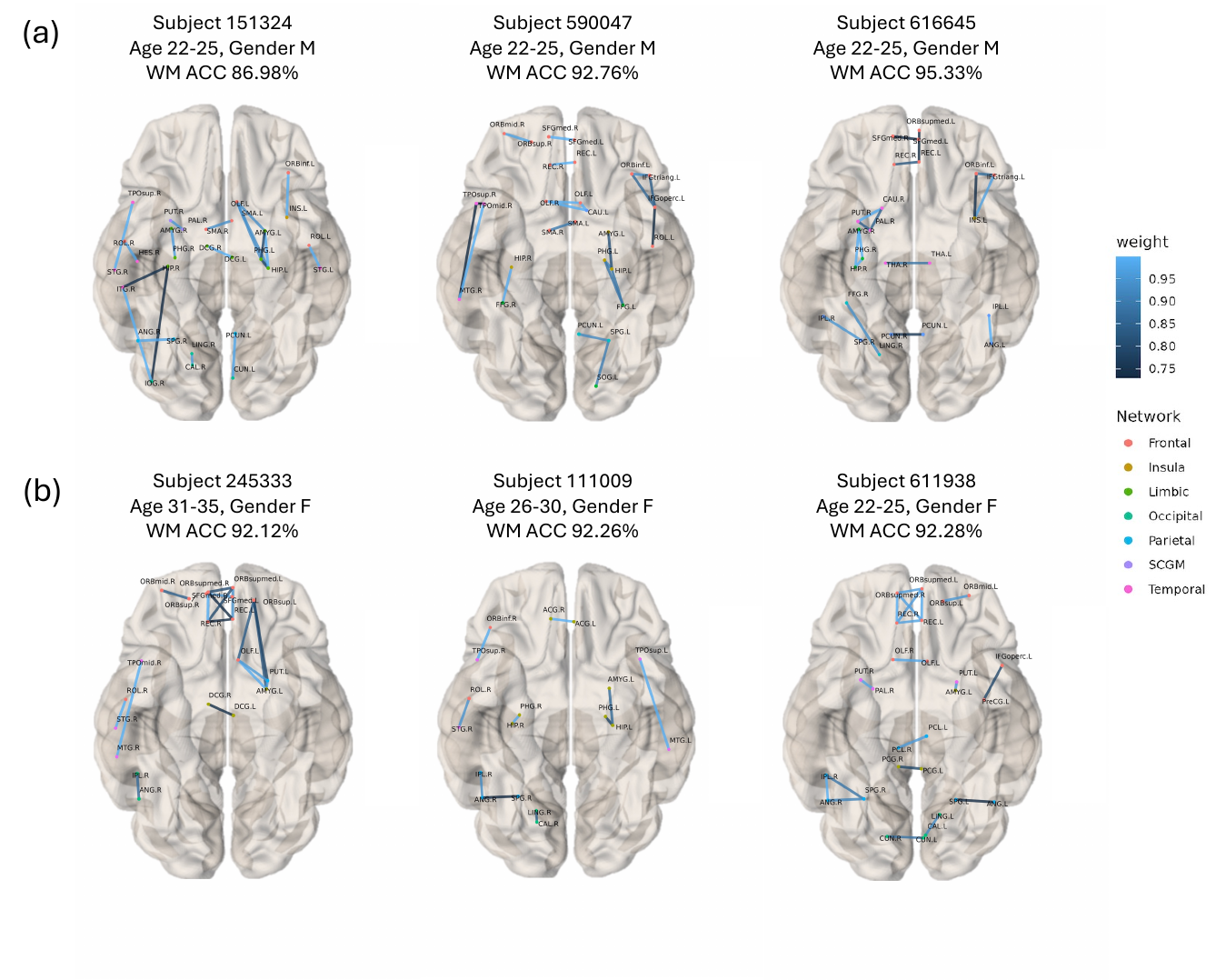}
    \caption{The brain coactivation networks between the AAL regions for different subjects, where connected edges represent prior correlations greater than $0.7$ between activation regions. (a) Networks for three subjects of the same age group and gender but with different working memory task accuracies. (b) Networks for three subjects from different age groups but with similar working memory task accuracies.}
    \label{fig:network_connection}
\end{figure}

We segment the brain into smaller regions using the Automated Anatomical Labeling (AAL) atlas \citep{TZOURIOMAZOYER2002273} and characterize brain coactivation between AAL regions based on prior correlations estimated by FLGP. Figure \ref{fig:network_connection} illustrates the coactivation networks across AAL regions for different individuals, revealing substantial heterogeneity in brain coactivation.
For subjects with varying working memory task accuracies (Figure \ref{fig:network_connection}a), both the network structure and connection strengths exhibit notable differences. Even among individuals with similar task accuracies (Figure \ref{fig:network_connection}b), variations in coactivation patterns and strengths persist. 
This heterogeneity among subjects with comparable task accuracies may stem from individual characteristics such as age, gender, or other unmeasured demographic factors. These findings underscore the variability in brain activity across individuals and emphasize the importance of domain-adaptive methods like FLGP, which can estimate subject-specific manifolds tailored to individual differences.

\begin{figure}
    \centering
    \includegraphics[width=0.9\linewidth]{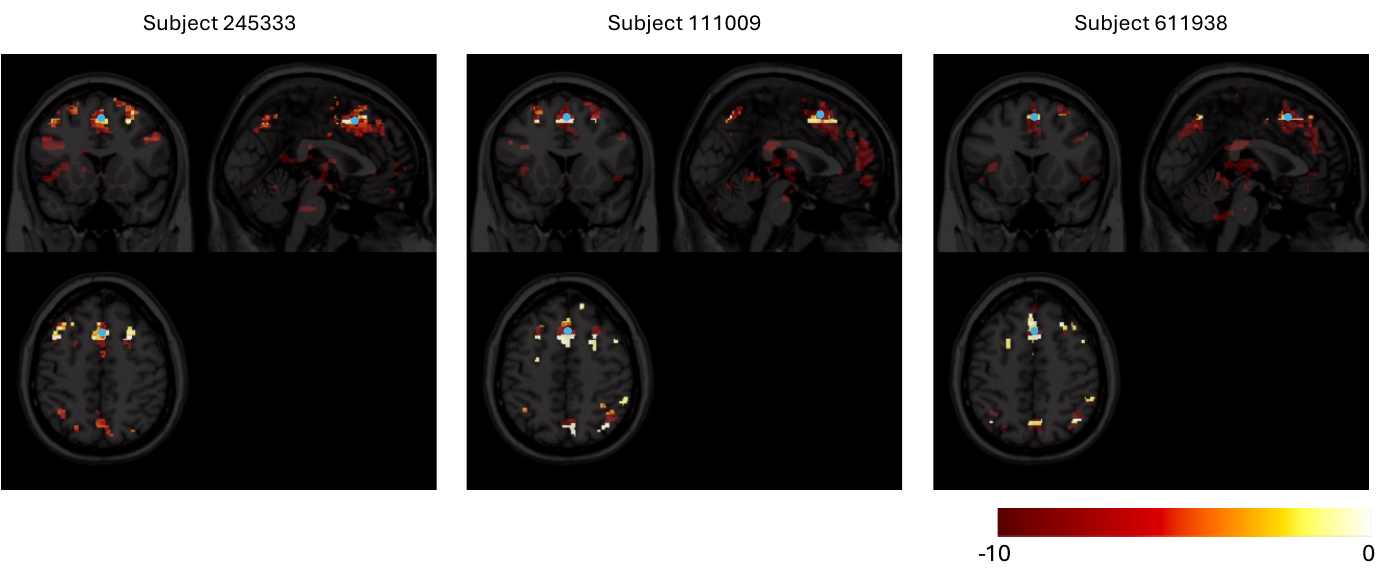}
    \caption{ Three-view heatmaps of log prior correlations between activated voxels (colored points) and the seed voxel (blue point), estimated by FLGP for the three subjects in Figure \ref{fig:network_connection}b.} 
    \label{fig:seed_voxel_cor_map}
\end{figure}

{ To investigate the pattern of brain activation, Figure \ref{fig:seed_voxel_cor_map} presents voxel-level log prior correlations estimated by FLGP between activated voxels and a seed voxel. The seed voxel is selected as the location with the highest average intensity within the common activation region. Intuitively, the negative log correlation can be interpreted as a measure of intrinsic distance. As shown in Figure \ref{fig:seed_voxel_cor_map}, some activated voxels that are close to the seed voxel in Euclidean space exhibit larger intrinsic distances than voxels that are spatially farther away. This phenomenon arises from the highly folded nature of brain anatomy, particularly in cortical regions \citep{power2011functional, mai2015atlas}. Two voxels that are Euclidean-close may lie on opposite sides of a cortical fold and thus be anatomically distant, while voxels that appear distant may reside on the same functional or anatomical tract. This anatomical complexity poses challenges for standard Euclidean metrics and underscores the importance of manifold learning methods in capturing meaningful brain structure.}

\section{Conclusion}

In this article, we introduce a scalable Bayesian inference method for heat kernel GPs on an unknown $d$-submanifold of $\mathbb{R}^p$. We develop a reduced-rank approximation for the transition matrix of a random walk on the point cloud and implement an efficient TSVD to avoid the computationally intensive eigendecomposition. These innovations significantly expedite the estimation of the heat kernel. Our proposed method demonstrates superior performance in highly non-Euclidean domain spaces compared to traditional Euclidean GPs and other intrinsic GPs.

{ We note that although the motivation behind domain-adaptive GPs is to incorporate the intrinsic geometry of the domain manifold, the resulting kernel is primarily estimated to optimize regression performance rather than to faithfully approximate the manifold. Specifically, in our graph-based approach, the geometric information of the manifold is encoded solely by the connectivity structure of the graph on the point cloud. When the squared exponential kernel serves as the base kernel, this connectivity pattern depends heavily on the bandwidth parameter $\epsilon$. Ideally, for a given distribution of covariates, there may exist an optimal bandwidth that best captures the geometry. However, in our algorithm and many other graph-based methods, the bandwidth is selected through maximizing the marginal likelihood tailored to a specific regression task. Consequently, the constructed graph may not accurately approximate the manifold. Nevertheless, as indicated by numerical experiments, our proposed method achieve good regression performance, even though geometric fidelity of the graph may be compromised.} 

Looking ahead, several interesting avenues for future research present themselves. Firstly, integrating our FLGP with known geometries is a promising direction. In many practical scenarios, such as \citet{GUINNESS2016143}, where the manifold $\mathcal{X}$ is given, leveraging this prior knowledge could further improve performance. Secondly, investigating the sample paths of heat kernel GPs, including their smoothness, is another area of interest. 
Thirdly, exploring the possibility of employing powerful graph neural networks as an alternative to graph Laplacian for extracting domain information is a cutting-edge direction for research.

\section{Code and supplementary materials}

The source package FLGP is accessible on the GitHub repository \url{https://github.com/junhuihe2000/FLGP}, and \texttt{R} codes of the examples in Section 5 are available on the GitHub repository \url{https://github.com/junhuihe2000/exampleFLGP}. The Supplementary Materials provide all the proofs, the implementation details of the Nystr\"om extension, evaluations of heat kernel approximation accuracy, sensitivity analyses of the predefined hyperparameters and the impact of sampling distributions. They also include a fast GP regression algorithm designed for large labeled observations.
\section*{Acknowledgements}

We sincerely thank the editor, the associate editor, and the referees for their comments to improve the quality of the paper.  Yang's research was partially supported by the National Natural Science Foundation of China grant (12271286, 11931001).

\bibliographystyle{rss}
\bibliography{refs/gp}

\begin{thebibliography}{46}
\expandafter\ifx\csname natexlab\endcsname\relax\def\natexlab#1{#1}\fi
\expandafter\ifx\csname url\endcsname\relax
  \def\url#1{\texttt{#1}}\fi
\expandafter\ifx\csname urlprefix\endcsname\relax\def\urlprefix{URL: }\fi

\bibitem[{Azangulov et~al.(2023)Azangulov, Smolensky, Terenin and Borovitskiy}]{azangulov2023stationarykernelsgaussianprocesses}
Azangulov, I., Smolensky, A., Terenin, A. and Borovitskiy, V. (2023) Stationary kernels and {G}aussian processes on {L}ie groups and their homogeneous spaces {I}: the compact case.
\newblock \textit{Preprint arXiv:2208.14960}.

\bibitem[{Baglama and Reichel(2005)}]{doi:10.1137/04060593X}
Baglama, J. and Reichel, L. (2005) Augmented implicitly restarted {L}anczos bidiagonalization methods.
\newblock \textit{SIAM Journal on Scientific Computing}, \textbf{27}, 19--42.

\bibitem[{Belkin and Niyogi(2001)}]{belkin2001laplacian}
Belkin, M. and Niyogi, P. (2001) {L}aplacian eigenmaps and spectral techniques for embedding and clustering.
\newblock In \textit{Advances in Neural Information Processing Systems}, vol.~14. Cambridge: MIT Press.

\bibitem[{Castillo et~al.(2014)Castillo, Kerkyacharian and Picard}]{castillo2014thomas}
Castillo, I., Kerkyacharian, G. and Picard, D. (2014) Thomas {B}ayes’ walk on manifolds.
\newblock \textit{Probability Theory and Related Fields}, \textbf{158}, 665--710.

\bibitem[{Coifman and Lafon(2006)}]{coifman2006diffusion}
Coifman, R.~R. and Lafon, S. (2006) Diffusion maps.
\newblock \textit{Applied and computational harmonic analysis}, \textbf{21}, 5--30.

\bibitem[{Diestel(2024)}]{diestel2024graph}
Diestel, R. (2024) \textit{Graph Theory}, vol. 173.
\newblock Heidelberg: Springer.

\bibitem[{Dunson et~al.(2021{\natexlab{a}})Dunson, Wu and Wu}]{10.1111/rssb.12486}
Dunson, D.~B., Wu, H.-T. and Wu, N. (2021{\natexlab{a}}) Graph based {G}aussian processes on restricted domains.
\newblock \textit{Journal of the Royal Statistical Society Series B: Statistical Methodology}, \textbf{84}, 414--439.

\bibitem[{Dunson et~al.(2021{\natexlab{b}})Dunson, Wu and Wu}]{DUNSON2021282}
--- (2021{\natexlab{b}}) Spectral convergence of graph {L}aplacian and heat kernel reconstruction in ${L}^\infty$ from random samples.
\newblock \textit{Applied and Computational Harmonic Analysis}, \textbf{55}, 282--336.

\bibitem[{Feragen et~al.(2015)Feragen, Lauze and Hauberg}]{Feragen_2015_CVPR}
Feragen, A., Lauze, F. and Hauberg, S. (2015) Geodesic exponential kernels: When curvature and linearity conflict.
\newblock In \textit{Proceedings of the IEEE Conference on Computer Vision and Pattern Recognition}, 3032--3042.

\bibitem[{Fichera et~al.(2023)Fichera, Borovitskiy, Krause and Billard}]{NEURIPS2023_d611d06e}
Fichera, B., Borovitskiy, S., Krause, A. and Billard, A.~G. (2023) Implicit manifold {G}aussian process regression.
\newblock In \textit{Advances in Neural Information Processing Systems}, vol.~36, 67701--67720. Curran Associates, Inc.

\bibitem[{Fischl et~al.(1999)Fischl, Sereno and Dale}]{fischl1999cortical}
Fischl, B., Sereno, M.~I. and Dale, A.~M. (1999) Cortical surface-based analysis {II}: Inflation, flattening, and a surface-based coordinate system.
\newblock \textit{Neuroimage}, \textbf{9}, 195--207.

\bibitem[{Garc{\'\i}a~Trillos et~al.(2020)Garc{\'\i}a~Trillos, Gerlach, Hein and Slep{\v{c}}ev}]{garcia2020error}
Garc{\'\i}a~Trillos, N., Gerlach, M., Hein, M. and Slep{\v{c}}ev, D. (2020) Error estimates for spectral convergence of the graph {L}aplacian on random geometric graphs toward the {L}aplace--{B}eltrami operator.
\newblock \textit{Foundations of Computational Mathematics}, \textbf{20}, 827--887.

\bibitem[{Genton(2001)}]{genton2001classes}
Genton, M.~G. (2001) Classes of kernels for machine learning: A statistics perspective.
\newblock \textit{Journal of Machine Learning Research}, \textbf{2}, 299--312.

\bibitem[{Greenbaum et~al.(2019)Greenbaum, Li and Overton}]{greenbaum2019firstorderperturbationtheoryeigenvalues}
Greenbaum, A., Li, R. and Overton, M.~L. (2019) First-order perturbation theory for eigenvalues and eigenvectors.
\newblock \textit{Preprint arXiv:1903.00785}.

\bibitem[{Grigor'yan(2006)}]{grigor2006heat}
Grigor'yan, A. (2006) Heat kernels on weighted manifolds and applications.
\newblock \textit{Contemporary Mathematics}, \textbf{398}, 93--191.

\bibitem[{Grigor'yan(2009)}]{grigoryan2009heat}
--- (2009) \textit{Heat Kernel and Analysis on Manifolds}, vol.~47.
\newblock Rhode Island: American Mathematical Society.

\bibitem[{Guinness and Fuentes(2016)}]{GUINNESS2016143}
Guinness, J. and Fuentes, M. (2016) Isotropic covariance functions on spheres: Some properties and modeling considerations.
\newblock \textit{Journal of Multivariate Analysis}, \textbf{143}, 143--152.

\bibitem[{Jonathan R.~Bradley and Wikle(2020)}]{doi:10.1080/01621459.2019.1677471}
Jonathan R.~Bradley, S. H.~H. and Wikle, C.~K. (2020) Bayesian hierarchical models with conjugate full-conditional distributions for dependent data from the natural exponential family.
\newblock \textit{Journal of the American Statistical Association}, \textbf{115}, 2037--2052.

\bibitem[{Kendall et~al.(2009)Kendall, Barden, Carne and Le}]{kendall2009shape}
Kendall, D.~G., Barden, D., Carne, T.~K. and Le, H. (2009) \textit{Shape and Shape Theory}.
\newblock Hoboken: John Wiley \& Sons.

\bibitem[{Kim et~al.(2021)Kim, Dryden, Le and Severn}]{https://doi.org/10.1111/rssb.12402}
Kim, K.-R., Dryden, I.~L., Le, H. and Severn, K.~E. (2021) Smoothing splines on {R}iemannian manifolds, with applications to 3{D} shape space.
\newblock \textit{Journal of the Royal Statistical Society: Series B (Statistical Methodology)}, \textbf{83}, 108--132.

\bibitem[{Lee(2012)}]{jmlee2012manifold}
Lee, J.~M. (2012) \textit{Introduction to Smooth Manifolds}.
\newblock New York: Springer.

\bibitem[{Lin et~al.(2019)Lin, Mu, Cheung and Dunson}]{lin2019extrinsic}
Lin, L., Mu, N., Cheung, P. and Dunson, D. (2019) Extrinsic {G}aussian processes for regression and classification on manifolds.
\newblock \textit{Bayesian Analysis}, \textbf{14}, 887--906.

\bibitem[{Liu et~al.(2010)Liu, He and Chang}]{liu2010large}
Liu, W., He, J. and Chang, S.-F. (2010) Large graph construction for scalable semi-supervised learning.
\newblock In \textit{Proceedings of the 27th International Conference on Machine Learning}, 679--686.

\bibitem[{Lu et~al.(2025)Lu, Kochunov, Chen, Huang, Hong and Chen}]{lu2025new}
Lu, T., Kochunov, P., Chen, C., Huang, H.-H., Hong, L.~E. and Chen, S. (2025) A new multiple imputation method for high-dimensional neuroimaging data.
\newblock \textit{Human Brain Mapping}, \textbf{46}, 70161.

\bibitem[{Mai et~al.(2015)Mai, Majtanik and Paxinos}]{mai2015atlas}
Mai, J.~K., Majtanik, M. and Paxinos, G. (2015) \textit{Atlas of the human brain}.
\newblock Academic Press.

\bibitem[{Mulugeta et~al.(2017)Mulugeta, Eckert, Vaden, Johnson and Lawson}]{mulugeta2017methods}
Mulugeta, G., Eckert, M.~A., Vaden, K.~I., Johnson, T.~D. and Lawson, A.~B. (2017) Methods for the analysis of missing data in {FMRI} studies.
\newblock \textit{Journal of Biometrics \& Biostatistics}, \textbf{8}, 335.

\bibitem[{Nesterov(2018)}]{nesterov2018lectures}
Nesterov, Y. (2018) \textit{Lectures on Convex Optimization}, vol. 137.
\newblock Switzerland: Springer.

\bibitem[{Niu et~al.(2019)Niu, Cheung, Lin, Dai, Lawrence and Dunson}]{niu2019intrinsic}
Niu, M., Cheung, P., Lin, L., Dai, Z., Lawrence, N. and Dunson, D. (2019) Intrinsic {G}aussian processes on complex constrained domains.
\newblock \textit{Journal of the Royal Statistical Society: Series B (Statistical Methodology)}, \textbf{81}, 603--627.

\bibitem[{Niu et~al.(2023)Niu, Dai, Cheung and Wang}]{niu2023intrinsic}
Niu, M., Dai, Z., Cheung, P. and Wang, Y. (2023) Intrinsic {G}aussian process on unknown manifolds with probabilistic metrics.
\newblock \textit{Journal of Machine Learning Research}, \textbf{24}, 1--42.

\bibitem[{Power et~al.(2011)Power, Cohen, Nelson, Wig, Barnes, Church, Vogel, Laumann, Miezin, Schlaggar et~al.}]{power2011functional}
Power, J.~D., Cohen, A.~L., Nelson, S.~M., Wig, G.~S., Barnes, K.~A., Church, J.~A., Vogel, A.~C., Laumann, T.~O., Miezin, F.~M., Schlaggar, B.~L. et~al. (2011) Functional network organization of the human brain.
\newblock \textit{Neuron}, \textbf{72}, 665--678.

\bibitem[{Quinonero-Candela and Rasmussen(2005)}]{quinonero2005unifying}
Quinonero-Candela, J. and Rasmussen, C.~E. (2005) A unifying view of sparse approximate {G}aussian process regression.
\newblock \textit{Journal of Machine Learning Research}, \textbf{6}, 1939--1959.

\bibitem[{Rasmussen and Williams(2006)}]{rasmussen2006gaussian}
Rasmussen, C.~E. and Williams, C. K.~I. (2006) \textit{Gaussian Processes for Machine Learning}.
\newblock Cambridge: MIT Press.

\bibitem[{Rosa et~al.(2023)Rosa, Borovitskiy, Terenin and Rousseau}]{rosa2023posterior}
Rosa, P., Borovitskiy, S., Terenin, A. and Rousseau, J. (2023) Posterior contraction rates for {M}at{\'e}rn {G}aussian processes on {R}iemannian manifolds.
\newblock \textit{Advances in Neural Information Processing Systems}, \textbf{36}, 34087--34121.

\bibitem[{Roweis and Saul(2000)}]{roweis2000nonlinear}
Roweis, S.~T. and Saul, L.~K. (2000) Nonlinear dimensionality reduction by locally linear embedding.
\newblock \textit{Science}, \textbf{290}, 2323--2326.

\bibitem[{Schweinberger and Stewart(2020)}]{10.1214/19-AOS1810}
Schweinberger, M. and Stewart, J. (2020) {Concentration and consistency results for canonical and curved exponential-family models of random graphs}.
\newblock \textit{The Annals of Statistics}, \textbf{48}, 374--396.

\bibitem[{Sculley(2010)}]{sculley2010web}
Sculley, D. (2010) Web-scale k-means clustering.
\newblock In \textit{Proceedings of the 19th International Conference on World Wide Web}, 1177--1178.

\bibitem[{Shao(2003)}]{shao2003mathematical}
Shao, J. (2003) \textit{Mathematical Statistics}.
\newblock New York: Springer.

\bibitem[{Tzourio-Mazoyer et~al.(2002)Tzourio-Mazoyer, Landeau, Papathanassiou, Crivello, Etard, Delcroix, Mazoyer and Joliot}]{TZOURIOMAZOYER2002273}
Tzourio-Mazoyer, N., Landeau, B., Papathanassiou, D., Crivello, F., Etard, O., Delcroix, N., Mazoyer, B. and Joliot, M. (2002) Automated anatomical labeling of activations in {SPM} using a macroscopic anatomical parcellation of the {MNI} {MRI} single-subject brain.
\newblock \textit{NeuroImage}, \textbf{15}, 273--289.

\bibitem[{van~der Vaart and van Zanten(2008)}]{10.1214/009053607000000613}
van~der Vaart, A.~W. and van Zanten, J.~H. (2008) {Rates of contraction of posterior distributions based on Gaussian process priors}.
\newblock \textit{The Annals of Statistics}, \textbf{36}, 1435--1463.

\bibitem[{van~der Vaart and van Zanten(2009)}]{10.1214/08-AOS678}
--- (2009) {Adaptive Bayesian estimation using a Gaussian random field with inverse Gamma bandwidth}.
\newblock \textit{The Annals of Statistics}, \textbf{37}, 2655--2675.

\bibitem[{{van Essen} et~al.(2013){van Essen}, Smith, Barch, Behrens, Yacoub and Ugurbil}]{van2013wu}
{van Essen}, D.~C., Smith, S.~M., Barch, D.~M., Behrens, T.~E., Yacoub, E. and Ugurbil, K. (2013) The {WU-Minn} {H}uman {C}onnectome {P}roject: An overview.
\newblock \textit{NeuroImage}, \textbf{80}, 62--79.

\bibitem[{Weinberger and Saul(2006)}]{weinberger2006unsupervised}
Weinberger, K.~Q. and Saul, L.~K. (2006) Unsupervised learning of image manifolds by semidefinite programming.
\newblock \textit{International Journal of Computer Vision}, \textbf{70}, 77--90.

\bibitem[{Wu and Wu(2018)}]{10.1214/17-AOS1676}
Wu, H.-T. and Wu, N. (2018) {Think globally, fit locally under the manifold setup: Asymptotic analysis of locally linear embedding}.
\newblock \textit{The Annals of Statistics}, \textbf{46}, 3805--3837.

\bibitem[{Yan et~al.(2009)Yan, Huang and Jordan}]{10.1145/1557019.1557118}
Yan, D., Huang, L. and Jordan, M.~I. (2009) Fast approximate spectral clustering.
\newblock In \textit{Proceedings of the 15th ACM SIGKDD International Conference on Knowledge Discovery and Data Mining}, 907–916.

\bibitem[{Yang and Dunson(2016)}]{10.1214/15-AOS1390}
Yang, Y. and Dunson, D.~B. (2016) {Bayesian manifold regression}.
\newblock \textit{The Annals of Statistics}, \textbf{44}, 876--905.

\bibitem[{Ye et~al.(2024)Ye, Niu, Cheung, Dai and Liu}]{ye2020heat}
Ye, K., Niu, M., Cheung, P., Dai, Z. and Liu, Y. (2024) Intrinsic {G}aussian processes on manifolds and their accelerations by symmetry.
\newblock \textit{Preprint arXiv:2006.14266}.

\end{thebibliography}

\end{document}